\documentclass[10pt,letterpaper]{article}
\usepackage[top=0.85in,left=2.75in,footskip=0.75in]{geometry}
\usepackage{amsmath,amssymb}
\usepackage{changepage}
\usepackage{textcomp,marvosym}
\usepackage{cite}
\usepackage{nameref,hyperref}
\usepackage[right]{lineno}
\usepackage{microtype}
\DisableLigatures[f]{encoding = *, family = * }
\usepackage[table]{xcolor}
\usepackage{array}

\usepackage[utf8]{inputenc}
\usepackage[T1]{fontenc}
\usepackage{caption}
\usepackage{multirow}
\usepackage{amsmath}
\usepackage{rotating,tabularx}
\usepackage{caption}
\usepackage{subcaption}
\usepackage{graphicx}

\newcolumntype{+}{!{\vrule width 2pt}}

\newlength\savedwidth

\raggedright
\usepackage[aboveskip=1pt,labelfont=bf,labelsep=period,justification=raggedright,singlelinecheck=off]{caption}

\makeatletter
\renewcommand{\@biblabel}[1]{\quad#1.}
\makeatother

\usepackage{lastpage,fancyhdr,graphicx}
\usepackage{epstopdf}
\fancyheadoffset[L]{2.25in}
\fancyfootoffset[L]{2.25in}
\begin{document}
\vspace*{0.2in}

\begin{flushleft}
{\Large
\textbf\newline{An Investigation of Containment Measures Against the COVID-19 Pandemic in Mainland China} 
}
\newline
\\
Ji Liu\textsuperscript{1\dag},
Xiakai Wang\textsuperscript{1\dag},
Haoyi Xiong\textsuperscript{1},
Jizhou Huang\textsuperscript{1},
Siyu Huang\textsuperscript{1},
Haozhe An\textsuperscript{1},
Dejing Dou\textsuperscript{1*},
Haifeng Wang\textsuperscript{1}
\\
\bigskip
\textbf{1} Baidu Inc., Beijing, 100085, China
\\
\bigskip

\dag These authors contributed equally to this work.



* doudejing@baidu.com

\end{flushleft}

\begin{abstract}

As the recent COVID-19 outbreak rapidly expands all over the world, various containment measures have been carried out to fight against the COVID-19 pandemic. In Mainland China, the containment measures consist of three types, i.e., Wuhan travel ban, intra-city quarantine and isolation, and inter-city travel restriction. In order to carry out the measures, local economy and information acquisition play an important role. 
In this paper, we investigate the correlation of local economy and the information acquisition on the execution of containment measures to fight against the COVID-19 pandemic in Mainland China.
First, we use a parsimonious model, i.e., SIR-X model, to estimate the parameters, which represent the execution of intra-city quarantine and isolation in major cities of Mainland China.
In order to understand the execution of intra-city quarantine and isolation, we analyze the correlation between the representative parameters including local economy, mobility, and information acquisition. To this end, we collect the data of Gross Domestic Product (GDP), the inflows from Wuhan and outflows, and the COVID-19 related search frequency from a widely-used Web mapping service, i.e., Baidu Maps, and Web search engine, i.e., Baidu Search Engine, in Mainland China.
Based on the analysis, we confirm the strong correlation between the local economy and the execution of information acquisition in major cities of Mainland China. We further evidence that, although the cities with high GDP per capita attracts bigger inflows from Wuhan, people are more likely to conduct the quarantine measure and to reduce going out to other cities. Finally, the correlation analysis using search data shows that well-informed individuals are likely to carry out containment measures.

\end{abstract}


\section{Introduction}

\begin{figure}[ht]
    \centering
    \includegraphics[width=0.6\textwidth]{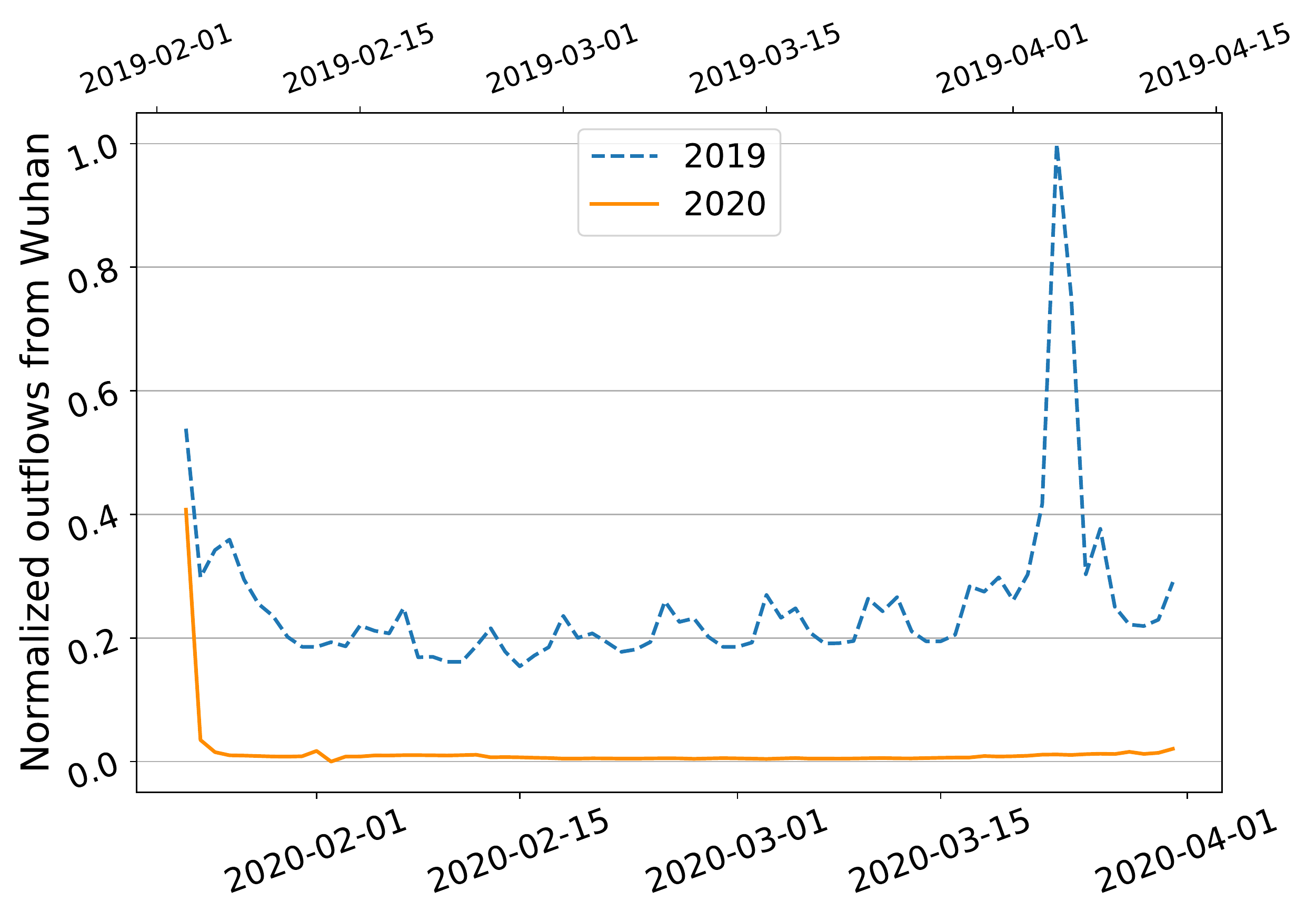}
    \caption{The comparisons of normalized outflows from Wuhan between 2020 and 2019 in Mainland China. We believe the peak on April 4, 2019 is due to the vacation of Qingming Festival.}
    \label{fig:wuhan_migration}
   \vspace{-5mm}
\end{figure}

Since December 2019, novel coronavirus COVID-19 has been identified and the outbreak has expanded rapidly throughout tremendous countries, e.g., China \cite{chinazzi2020effect}, United States \cite{JH2020}, European countries \cite{Deslandes2020}, etc.  In China, the number of confirmed cases increased from 571 on January 23, 2020 to 84,388 on May 1, 2020 and saturated around 84.5 thousand. The COVID-19 has become a global emergency and is currently spreading throughout the whole world~\cite{novel2020epidemiological,sohrabi2020world}.  In order to deal with the rapid outbreak of the COVID-19 pandemic in Mainland China, a range of containment measures have been put in place by Chinese authorities~\cite{kraemer2020effect,tian2020investigation,ferretti2020quantifying}. Similar containment measures have been adopted in major countries all over the world \cite{Engle2020,Remuzzi2020} etc.

\begin{figure}[ht]
    \centering
    \includegraphics[width=0.6\textwidth]{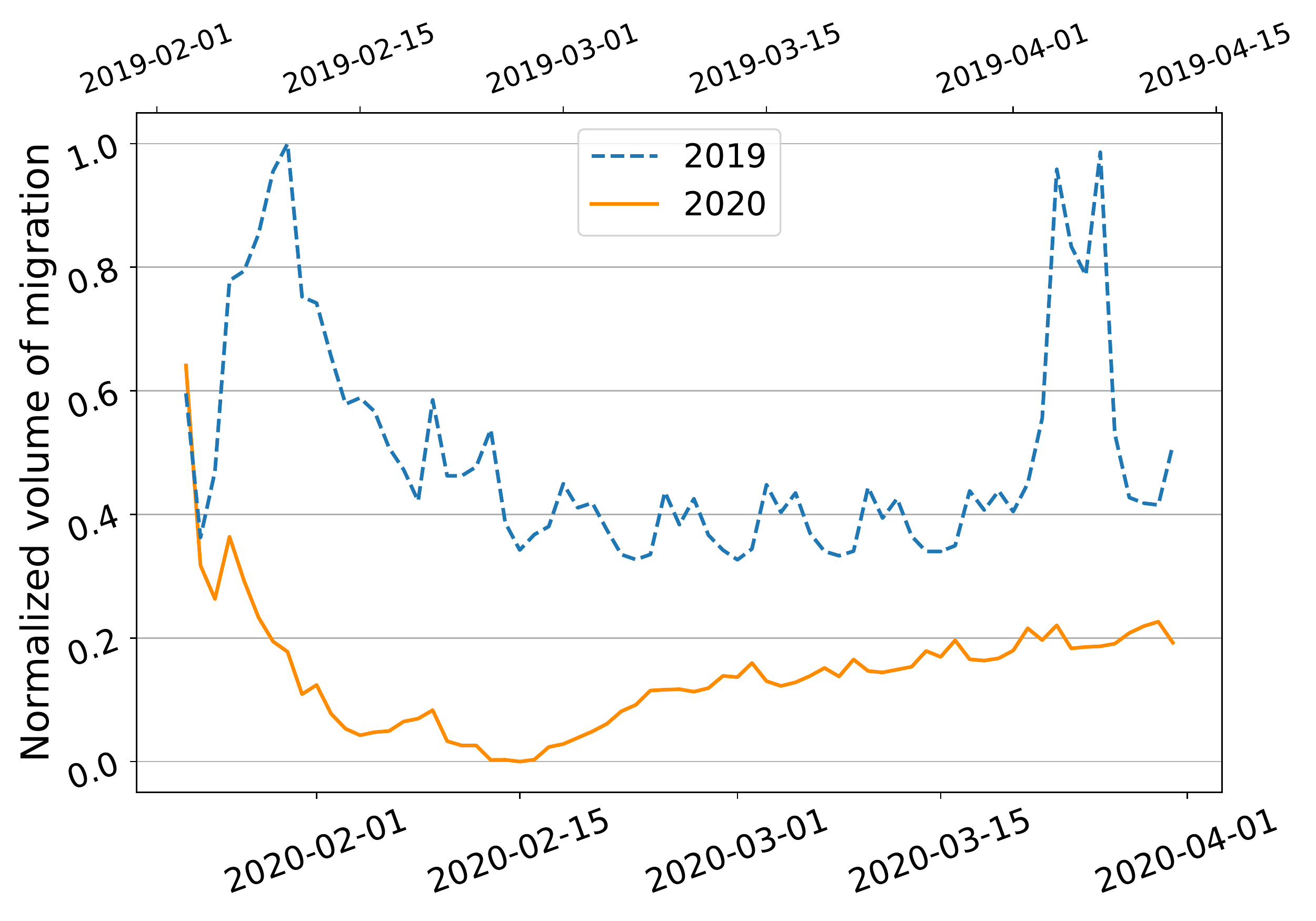}
    \caption{The comparisons of normalized volumes of the volume of migration between 2020 and 2019 in Mainland China. We believe the peaks between April 4, 2019 and April 7, 2019 are due to the vacation of Qingming Festival.} 
    \label{fig:china_migration}
   \vspace{-5mm}
\end{figure}

In Mainland China, the containment measures consist of intra-city and inter-city measures.
As intra-city measures, suspected and confirmed cases have been quarantined in hospitals or monitored self-quarantine at home \cite{Maier2020}, which is denoted the ``quarantine'' measure in the paper.
The authorities also encouraged citizens to stay-at-home and discouraged mass gatherings closed schools \cite{Chen2020}. In addition, Wuhan city travel ban was adopted, i.e., all transport was prohibited in and out of Wuhan city from 10:00 a.m. on 23 January 2020, which incurred a significant reduction of the outflows from Wuhan as shown in \ref{fig:wuhan_migration}.  As shown in Figure \ref{fig:china_migration}, the national spring vacation has been prolonged and inter-city travel has been discouraged to reduce massive human migration across cities in order to reduce infection.

Mobile applications, e.g., Baidu Migration \footnote{Baidu Migration - http://qianxi.baidu.com/}  and search engines, e.g., Baidu \footnote{Baidu - https://www.baidu.com/}, can be easily used to achieve information acquisition for citizens to keep informed during the outbreak of COVID-19.
There are a great number of studies~\cite{chinazzi2020effect,ren2020fear,yang2020modified,li2020novel,chen2020covid,zhang2020challenges,chen2020correlation,zhong2020correlation} that demonstrated the feasibility to leverage mobile applications for information acquisition.
As a result, the history search records can reflect the information acquisition status and the history statistical migration data can be used to analyze the status of the COVID-19 outbreak.
As well-informed individuals are likely to travel less when there is COVID-19 \cite{Xiong2020}, it is interesting to analyze the correlation between information acquisition and the execution of the containment measures. In addition, the correlation between the local economy and the information acquisition can be used to reveal how people in different economy situations react to the COVID-19 pandemic.

In this work, we aim at using a parsimonious model, i.e., SIR-X model, and Markov Chain Monte Carlo (MCMC) \cite{Gilks2005} methods to estimate the parameters of the execution of intra-city containment measures in major cities of Mainland China. 
Then, we analyze the correlation among different random variables, i.e., 
information acquisition status (COVID-19-related search frequency), local economy situation (GDP per capita) and the parameters in the SIR-X model, in order to understand the relationship among economy, information acquisition and the execution of containment measures. More specifically, we would like to investigate following problems:

\begin{figure*}[ht]
\centering
\begin{subfigure}{.32\textwidth}
  \centering
  \includegraphics[width=.99\linewidth]{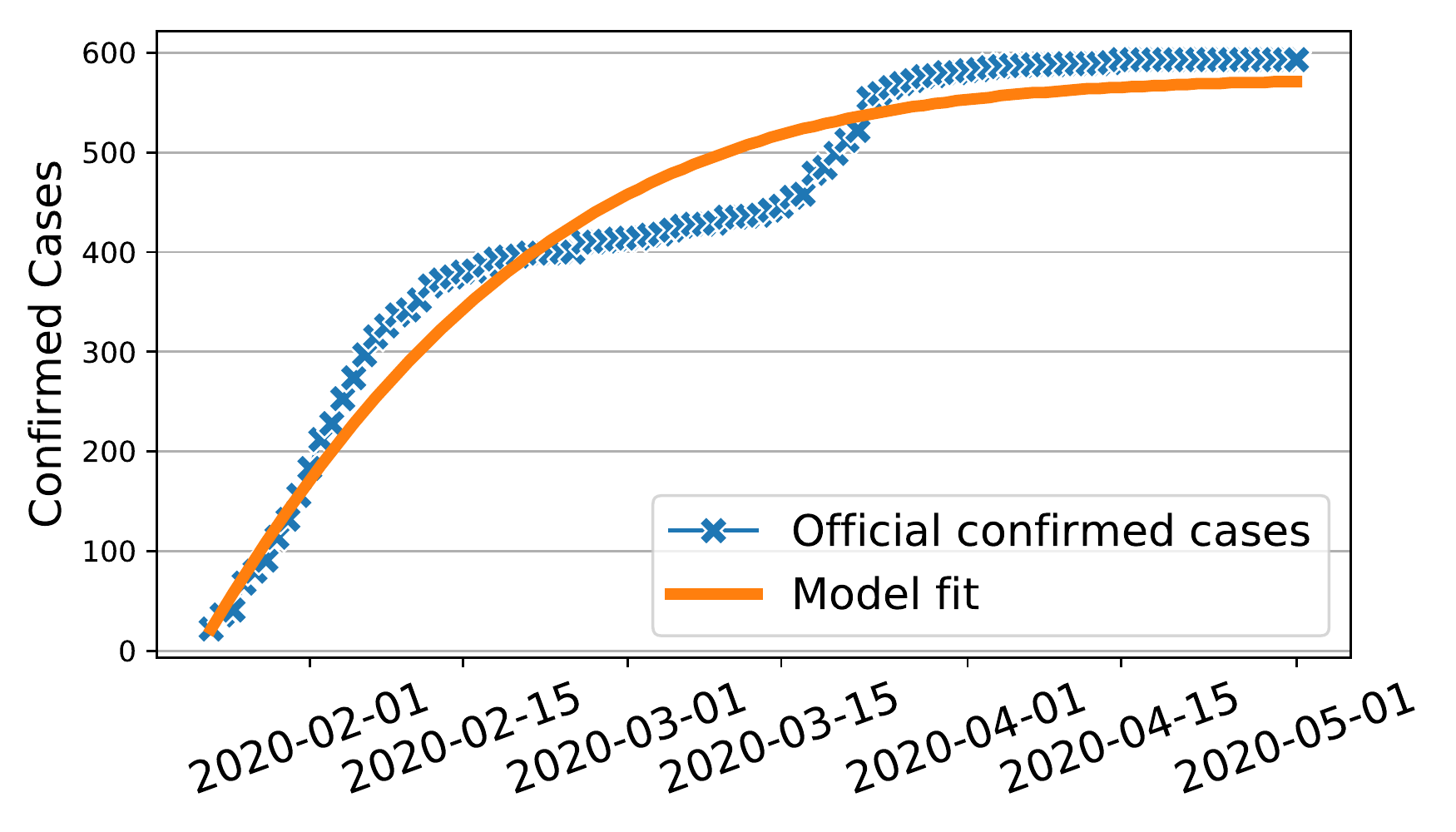}  
  \caption{Number of confirmed cases in Beijing.}
  \label{fig:sub-beijing}
\end{subfigure}
\begin{subfigure}{.32\textwidth}
  \centering
  \includegraphics[width=.9\linewidth]{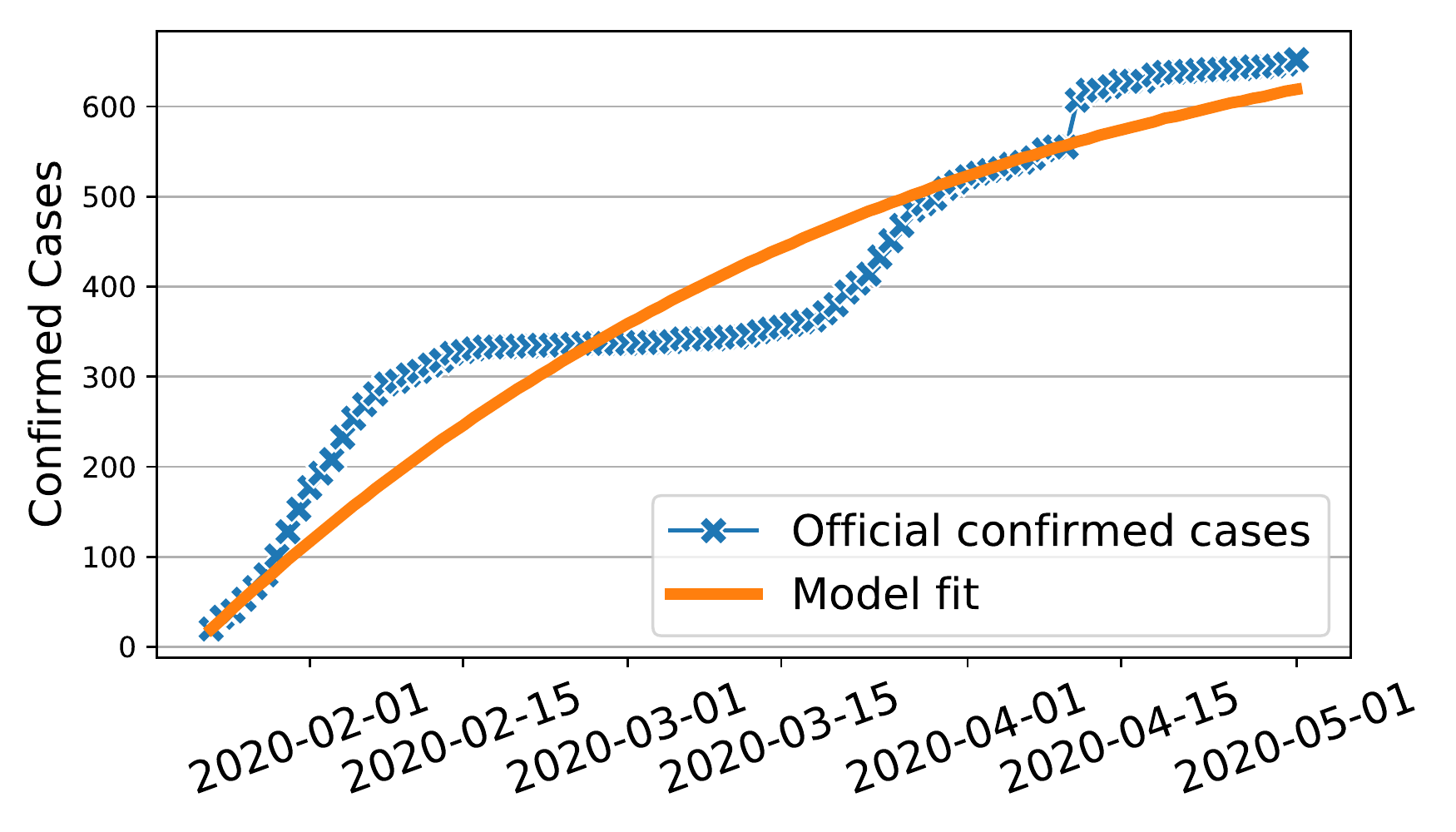}  
  \caption{Number of confirmed cases in Shanghai.}
  \label{fig:sub-shanghai}
\end{subfigure}
\begin{subfigure}{.32\textwidth}
  \centering
  \includegraphics[width=.9\linewidth]{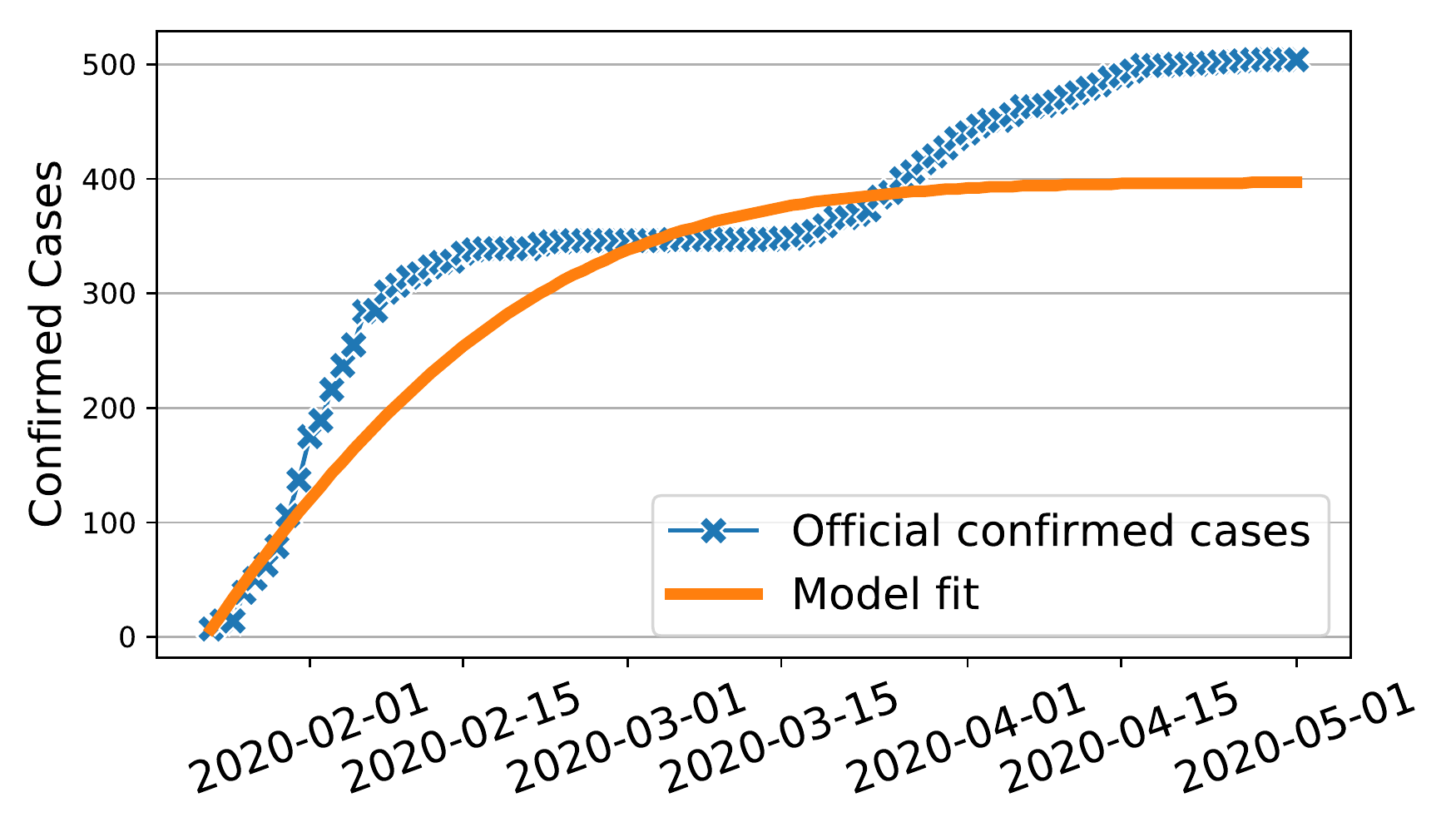}  
  \caption{Number of confirmed cases in Guangzhou.}
  \label{fig:sub-guangzhou}
\end{subfigure}
\\
\begin{subfigure}{.32\textwidth}
  \centering
  \includegraphics[width=.9\linewidth]{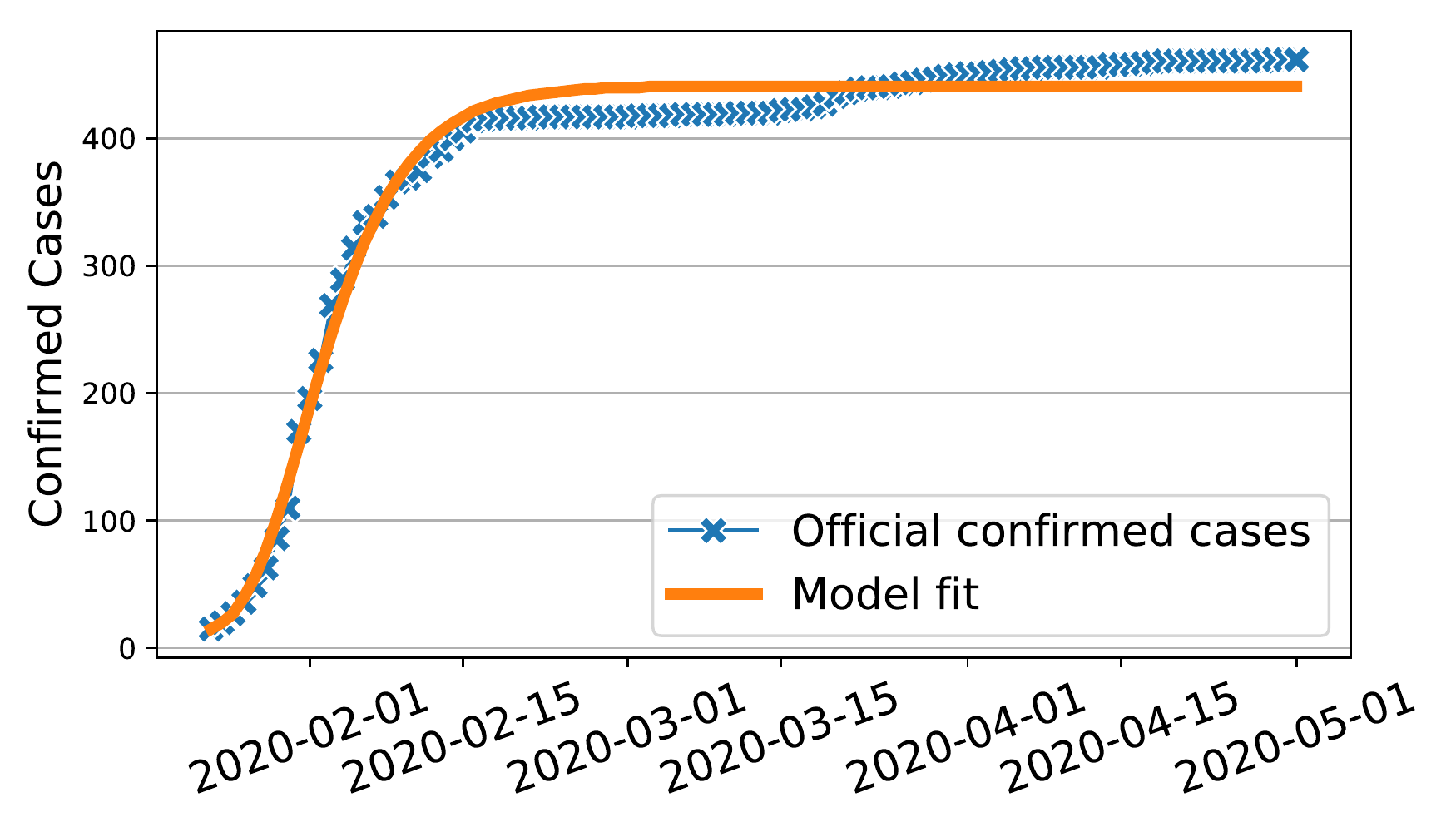}  
  \caption{Number of confirmed cases in Shenzhen.}
  \label{fig:sub-shenzhen}
\end{subfigure}
\begin{subfigure}{.32\textwidth}
  \centering
  \includegraphics[width=.9\linewidth]{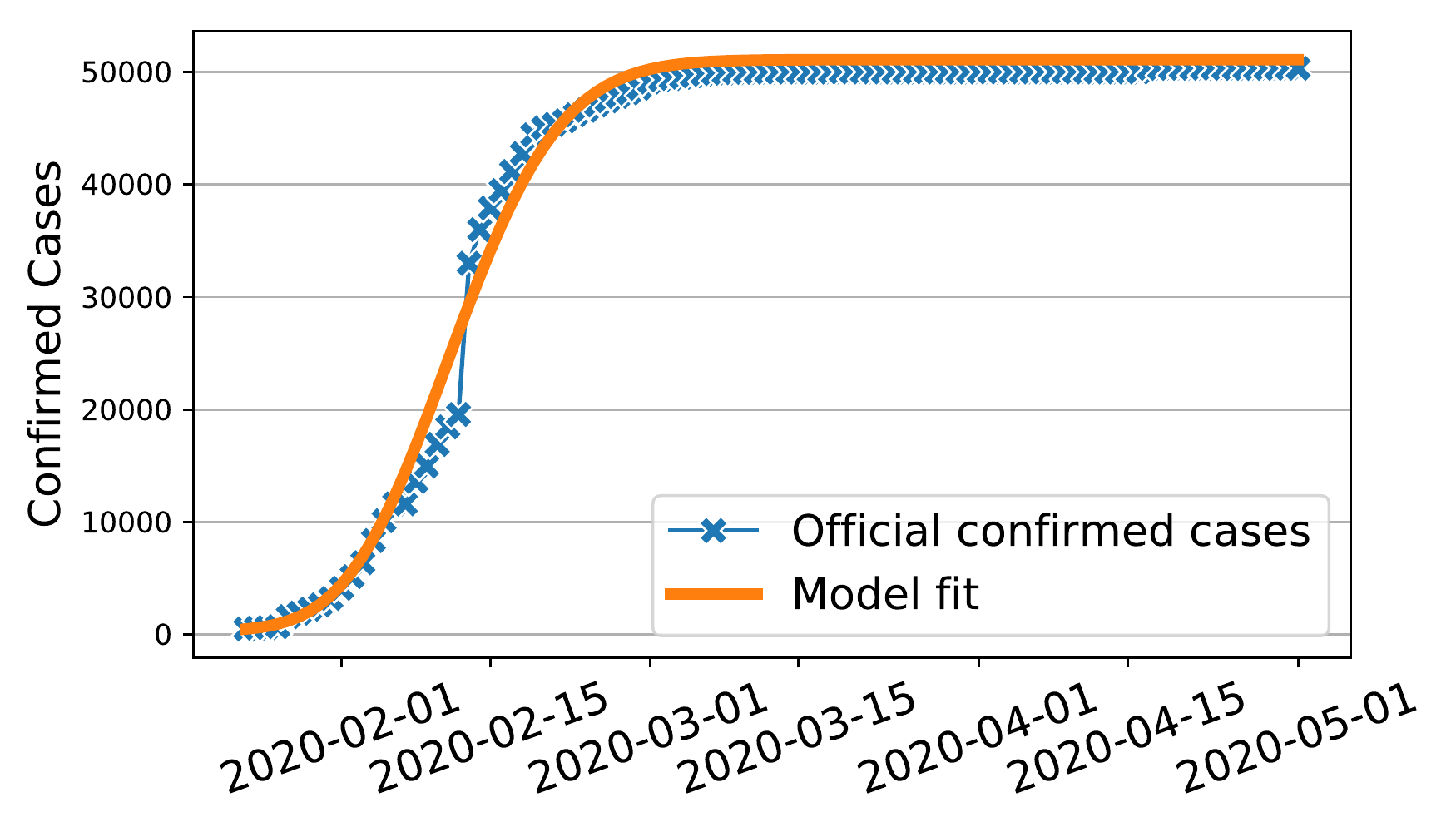}  
  \caption{Number of confirmed cases in Wuhan.}
  \label{fig:sub-wuhan}
\end{subfigure}
\begin{subfigure}{.32\textwidth}
  \centering
  \includegraphics[width=.9\linewidth]{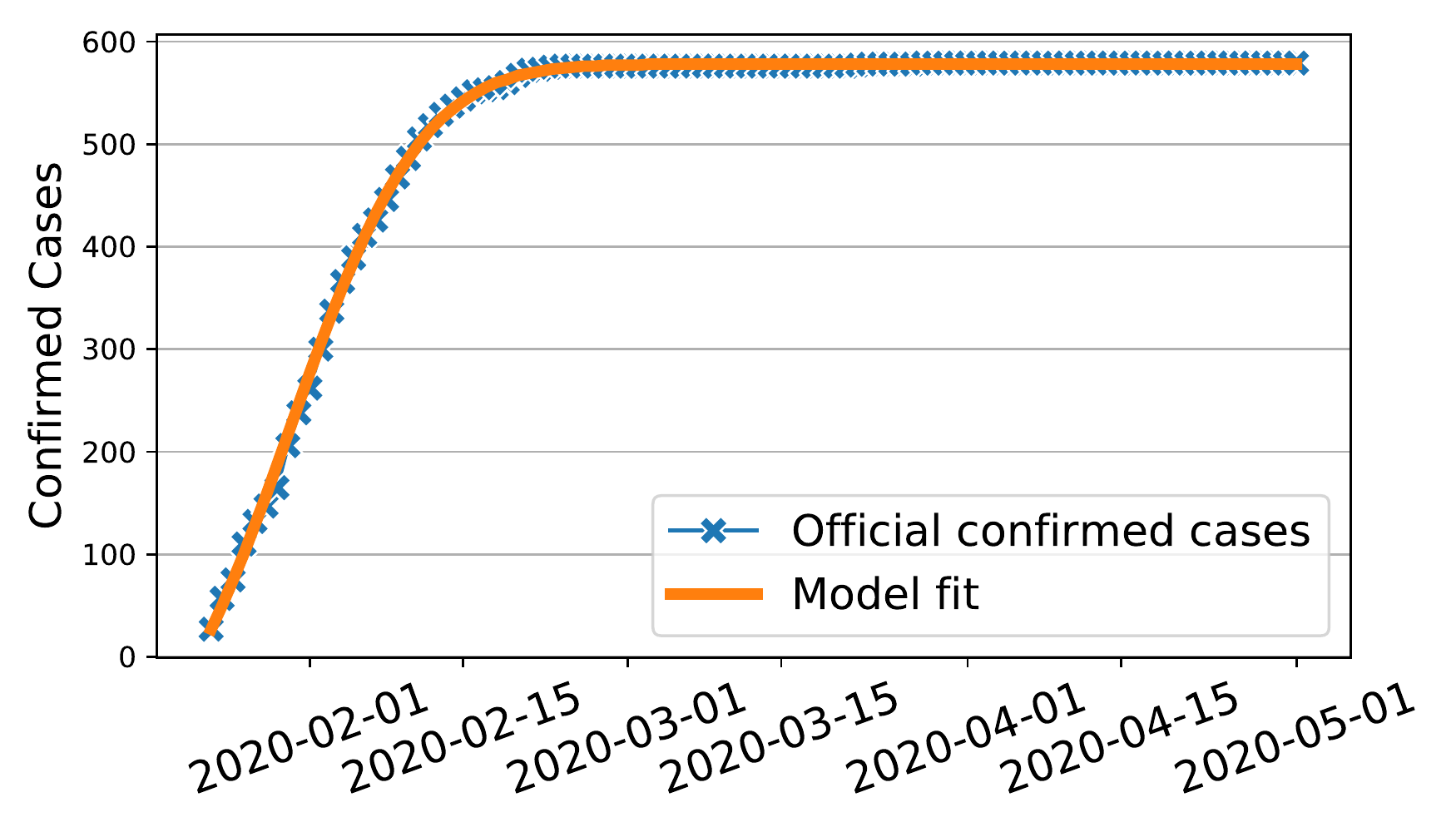}  
  \caption{Number of confirmed cases in Chongqing.}
  \label{fig:sub-chongqing}
\end{subfigure}
\caption{Comparison between official number of confirmed cases and fitting number from the SIR-X model on May 1, 2020 in Beijing, Shanghai, Guangzhou, Shenzhen, Wuhan and Chongqing.}
\label{fig:confirmedCases}
\vspace{-5mm}
\end{figure*}

\begin{itemize}

    \item \emph{How to construct a model to capture the confirmed cases?} This research issue has been studied~\cite{Maier2020,tian2020investigation} at province scale using the migration scale index released by Baidu Migration Open Data, where the index is calculated based on the past historical statistical records from a widely-used Web mapping service, i.e., Baidu Maps \footnote{Baidu  Maps - https://map.baidu.com/}. We propose to construct an SIR-X model based on the confirmed cases at city scale using a Markov Chain Monte Carlo (MCMC) method. We provide solid results using exact figures for major Chinese cities.
    
    \item \emph{To what degree does the local economy affect the pandemic outbreaks of COVID-19 and the execution of containment measures in major cities of China?} The impact of the COVID-19 pandemic on economy has been studied~\cite{huang2020quantifying,Atkeson2020} while the correlations between local economy and the outbreaks of the COVID-19 pandemic or the execution of containment measures are not analyzed. 
    We hypothesized that the outflows from Wuhan trend to go to the cities where GDP per capital is high. We further hypothesized that there would be more initial confirmed cases as more infected people arrived at these cities. 
    In addition, we hypothesized that the people in the cities where GDP per capital is high tend to perform more information acquisition activities through voluntary COVID-19-related search in order to be well informed on the situation of the COVID-19 pandemic.
    We analyze the correlations based on the estimated parameters of SIR-X and the statistical data from Baidu Maps and provide solid results for major Chinese cities.
    
    \item \emph{To what degree does the information acquisition affect the execution of containment measures in major cities of China?} Strong positive correlation between the pandemic outbreaks of COVID-19 the information acquisition has been reported in~\cite{Xiong2020} while the correlation between the information acquisition and the execution of containment measures are not analyzed. 
    As reported in \cite{Xiong2020} and was seen in the collective responses to the emergencies~\cite{gao2014quantifying} and panics~\cite{ren2020fear}, people voluntarily acquire information more frequently when the pandemic situations become worse in their cities.
    We hypothesized the well-informed people tend to apply the containment measures more strictly in order to avoid the risk to be infected and the risk to make the situation worse.
    We analyze this correlation based on the estimated parameters of SIR-X and the statistical data from Baidu Maps and Baidu Search Engine, and provide solid results for major Chinese cities as well.

\end{itemize}

Different from existing research \cite{Maier2020,tian2020investigation,kraemer2020effect,chinazzi2020effect}, we particularly analyze the correlation between local economy strength and the COVID-19-related search frequency with the city population size (a controlling variable) removed in order to avoid the impact of the scale of city. Compared to the existing work \cite{Xiong2020,huang2020quantifying}, we analyze the correlations not only based on the data from Baidu Maps and Baidu Search Engine but also based on the combination of an SIR-X model and MCMC methods.

\section{Modeling epidemic spread with containment measures}
\label{sec:modeling}

In this section, we first present the existing models to capture COVID-19. Then, we propose using SIR-X and MCMC to construct the model. Afterwards, we present the comparison between the official number and the model fitting number of accumulated confirmed cases.

Susceptible Infectious Recovered (SIR) model \cite{Toda2020,Simha2020} and Susceptible Exposed Infectious Recovered (SEIR) model \cite{Peng2020, Hou2020,Tang2020} are largely adopted to characterize the outbreak of COVID-19 epidemic.
However, the containment measures cannot be described in the standard SIR or SEIR model.
A modified SEIR model \cite{Yang2020} is proposed with the consideration of mobility while it is still not able to infer the execution of containment measures.
A Long-Short-Term-Memory (LSTM) \cite{Yang2020} model is proposed to project the number of accumulated confirmed cases, which is not able to describe the execution of containment measures either. 

In order to characterize the outbreak of COVID-19 epidemic with containment measures at city level, we exploit the SIR-X model \cite{Maier2020}.
The SIR-X model is a modified SIR model, which takes the containment measures into consideration. 
We have the same assumptions and use the same representative parameters as those in \cite{Maier2020}. We assume that there are public containment efforts, e.g., stay-at-home, reduced interaction with other people, which is referred as `containment' and represented by a variable $\kappa_0$.
In addition, we assume that infected individuals are quarantined, which is referred as `quarantine' and represented by a variable $\kappa$.
We use $\alpha$ to represent the infection speed of an infected individual and $\beta^{-1}$ to represent the average time an infected individual remains infectious before recovery or removal.
Then, the SIR-X model is expressed by the following differential equations:

\begin{equation}
\label{eq:sirx}
\begin{aligned}
    \partial_t S &= -\alpha SI - \kappa_0 S \\
    \partial_t I &= \alpha SI - \beta I - \kappa_0 I - \kappa I \\
    \partial_t R &= \beta I + \kappa_0 S \\
    \partial_t X &= (\kappa + \kappa_0)I \\
\end{aligned}
\end{equation}

Instead of fixing the same parameters ($\alpha$ and $\beta$) for each province in \cite{Maier2020}, we estimated the parameters using a MCMC \cite{Gilks2005} method, inspired by \cite{Wu2020}. 
In the model, $I_0$ represents the number of initial infected individuals. 
The basic reproduction number $R_0$ represents the average number of secondary infections an infected will cause before he or she recovers or is removed \cite{Maier2020}. 
The reproduction number can be calculated as: $R_0 = \dfrac{\alpha}{\beta + \kappa + \kappa_0}$.
We use $R_{0, free}$ to represent the reproduction number without containment or quarantine measures. 
As high temperature and high humidity significantly reduce the transmission of COVID-19 cite{Wang2020}, $R_{0, free}$ and $\beta$ may be different for different cities because of the diversity of local environments.
Thus, we use the MCMC \cite{Gilks2005} method to estimate the distribution of the parameters, i.e., $\alpha$, $\beta$, $\kappa$, $\kappa_0$, $I_0$, while the other parameters are fixed ($S_0$ is the population in the city, $R'_0$ is fixed as 0 and $X_0$ is the number of initial confirmed cases) at the beginning (January 23, 2020) with $S_0$, $R'_0$, $X_0$ and $I_0$ representing the initial values of $S$, $I$, $R$ and $X$. 

\begin{figure*}[ht]
\centering
\begin{subfigure}{.32\textwidth}
  \centering
  \includegraphics[width=.9\linewidth]{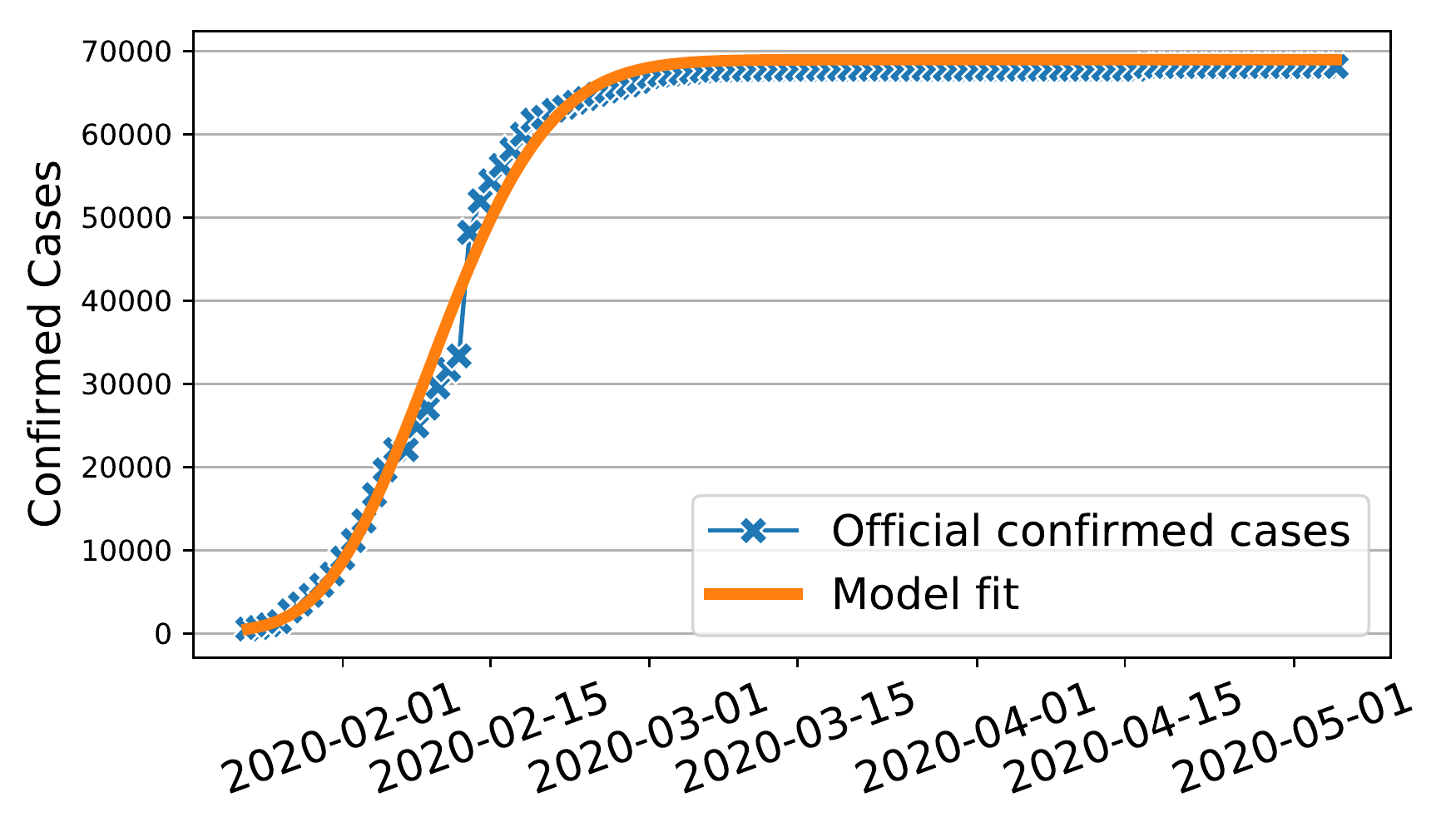}  
  \caption{Aggregated number of confirmed cases in Hubei.}
  \label{fig:sub-Hubei}
\end{subfigure}
\begin{subfigure}{.32\textwidth}
  \centering
  \includegraphics[width=.9\linewidth]{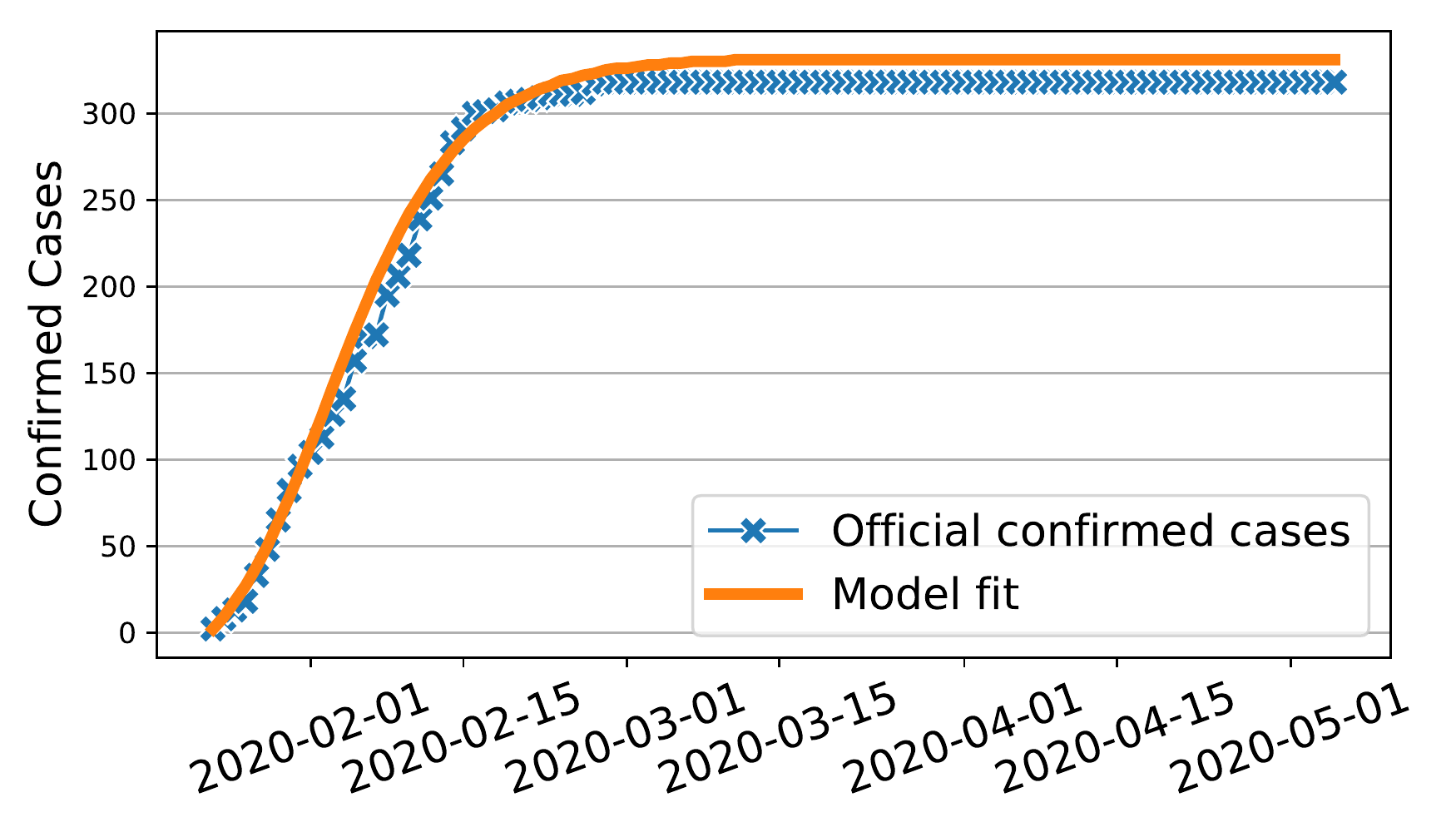}  
  \caption{Aggregated number of confirmed cases in Hebei.}
  \label{fig:sub-Hebei}
\end{subfigure}
\begin{subfigure}{.32\textwidth}
  \centering
  \includegraphics[width=.9\linewidth]{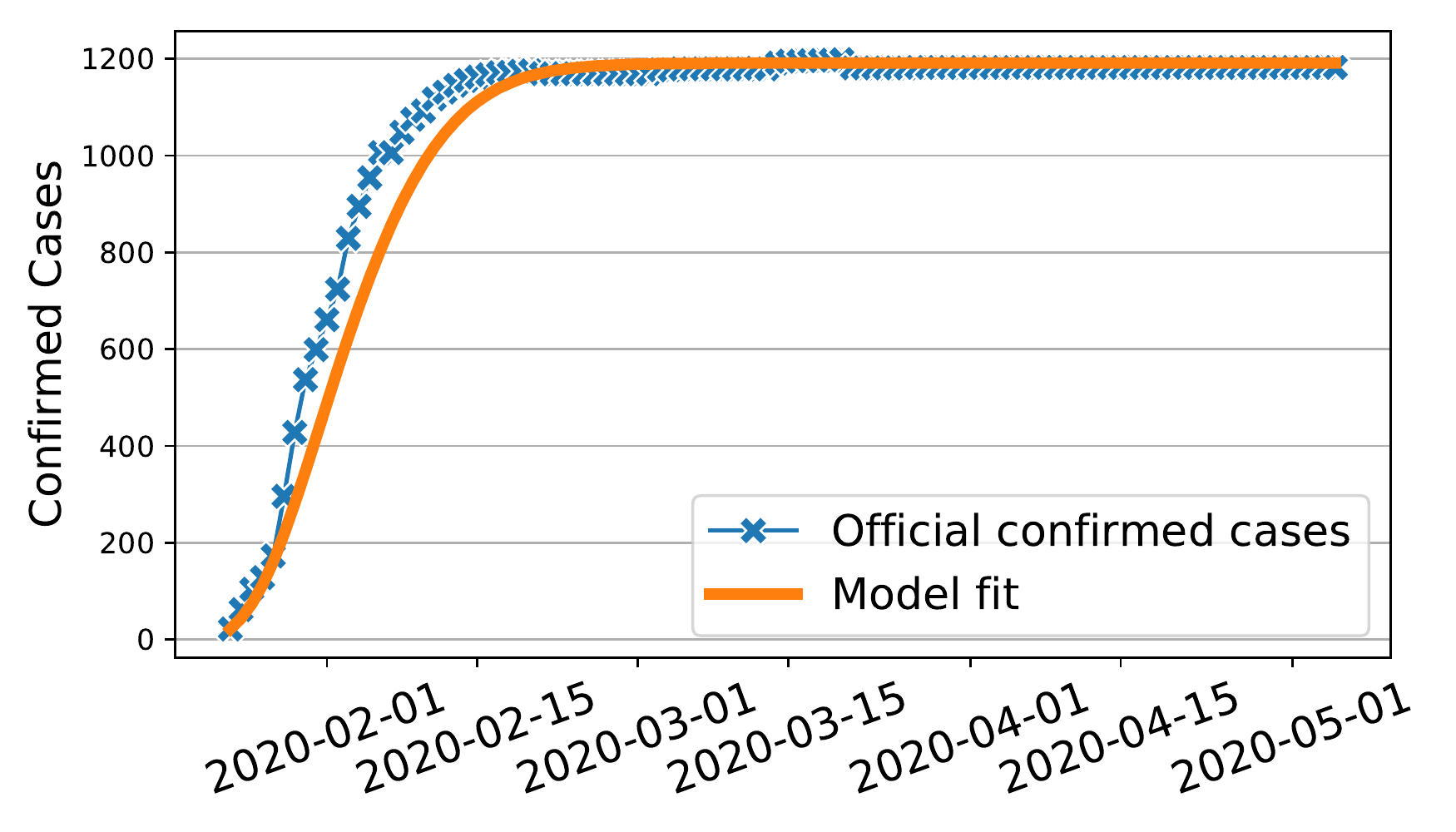}  
  \caption{Aggregated number of confirmed cases in \\ Zhejiang.}
  \label{fig:sub-Zhejiang}
\end{subfigure}
\\
\begin{subfigure}{.32\textwidth}
  \centering
  \includegraphics[width=.9\linewidth]{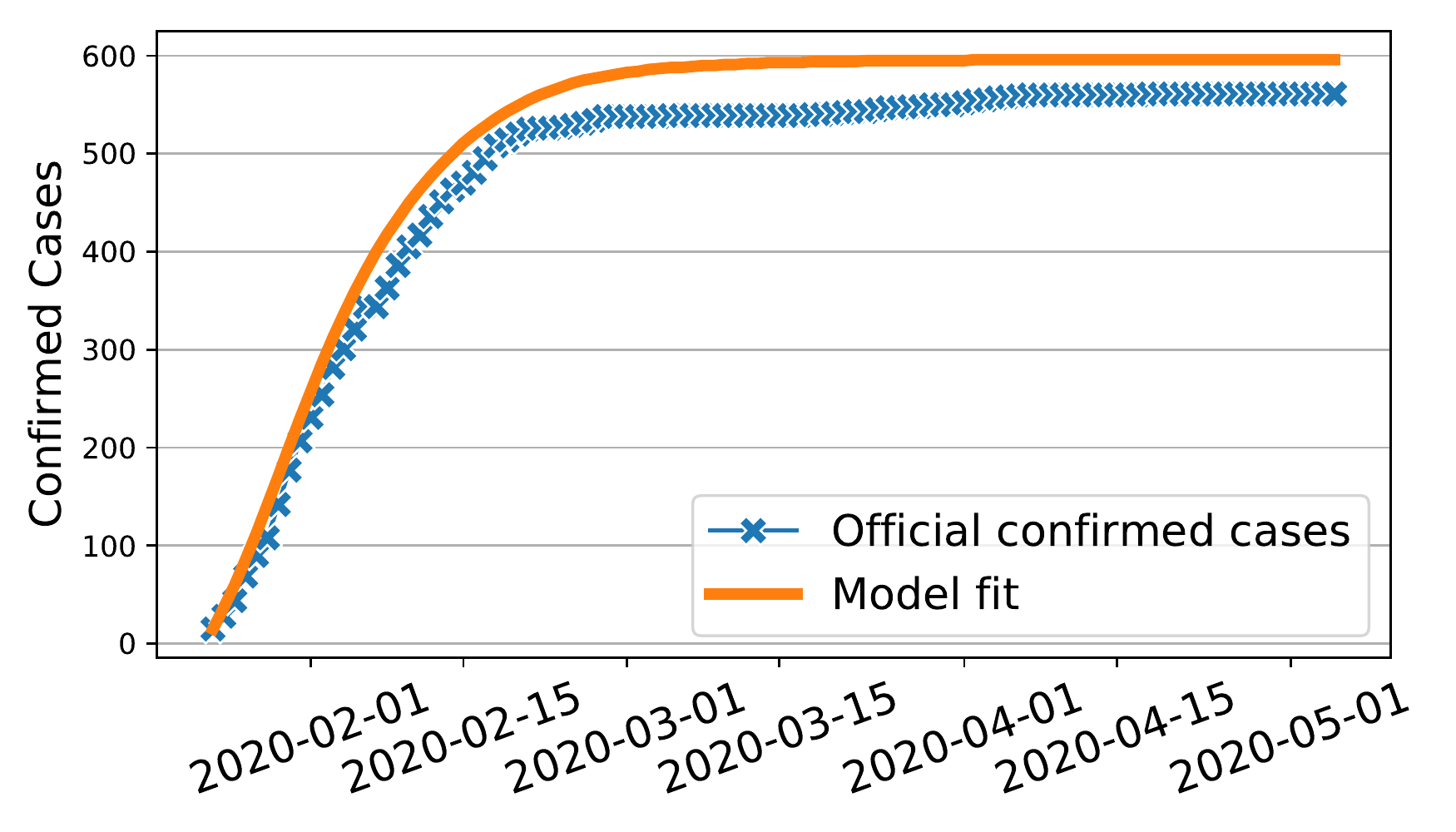}  
  \caption{Aggregated number of confirmed cases in Sichuan.}
  \label{fig:sub-Sichuan}
\end{subfigure}
\begin{subfigure}{.32\textwidth}
  \centering
  \includegraphics[width=.9\linewidth]{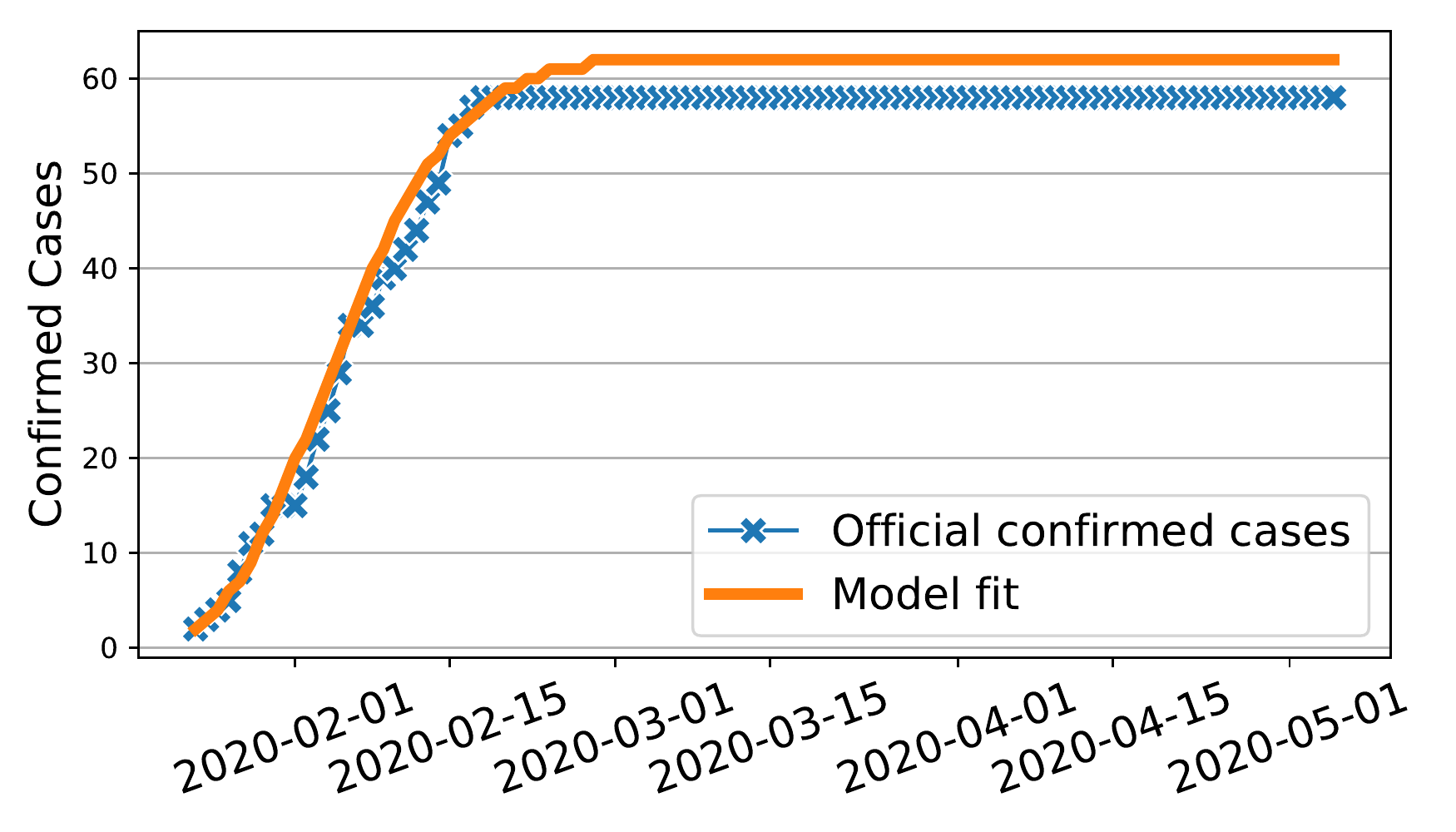}  
  \caption{Aggregated number of confirmed cases in \\ Xingjiang.}
  \label{fig:sub-Xingjiang}
\end{subfigure}
\begin{subfigure}{.32\textwidth}
  \centering
  \includegraphics[width=.9\linewidth]{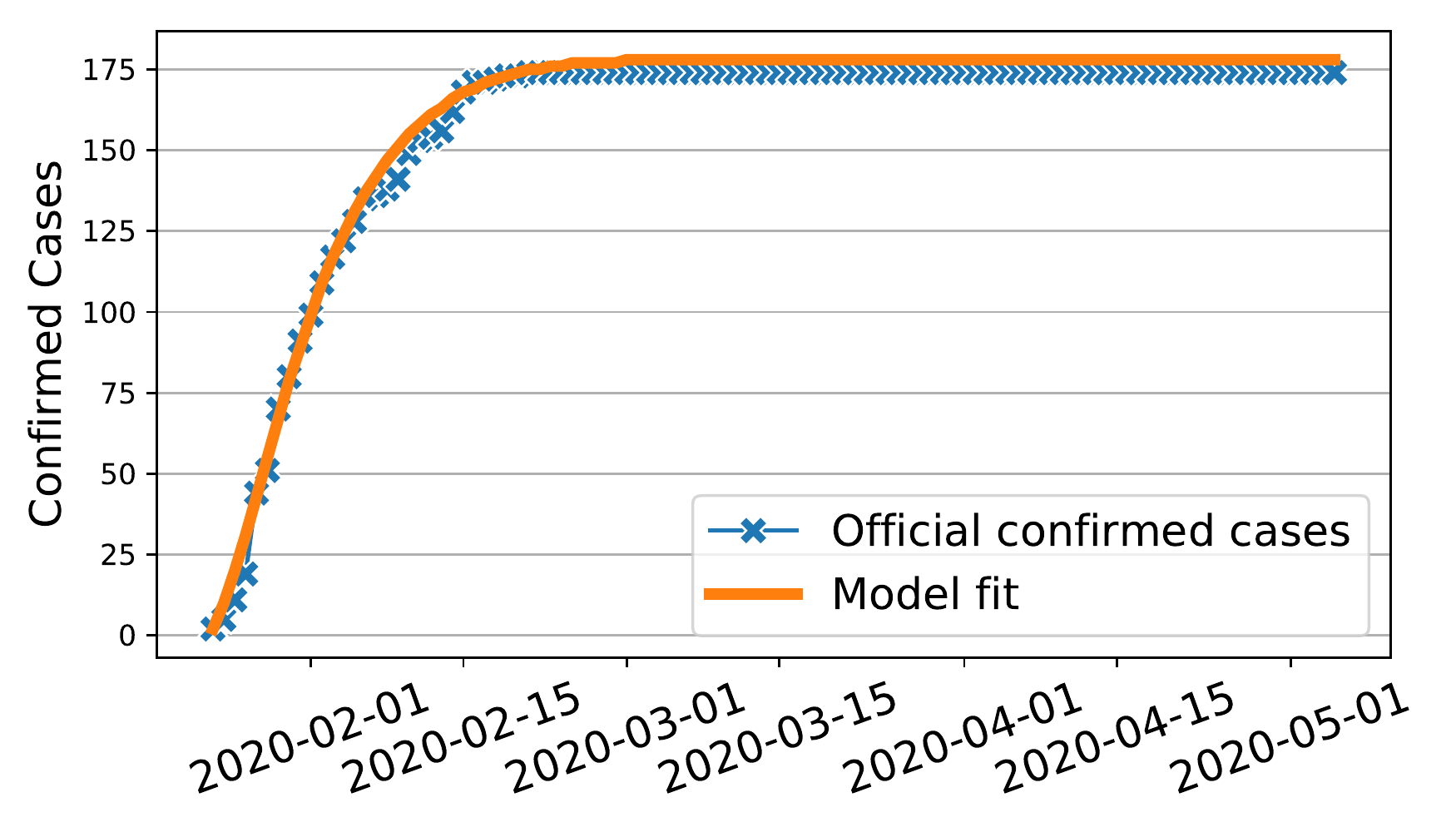}  
  \caption{Aggregated number of confirmed cases in Yunnan.}
  \label{fig:sub-Yunnan}
\end{subfigure}
\caption{Comparison between official number of confirmed cases and fitting number from the SIR-X model on May 1, 2020 in Hubei, Hebei, Zhejiang, Sichuan, Xingjiang and Yunnan.}
\vspace{-5mm}
\end{figure*}

\begin{figure}[ht]
    \centering
    \includegraphics[width=0.6\textwidth]{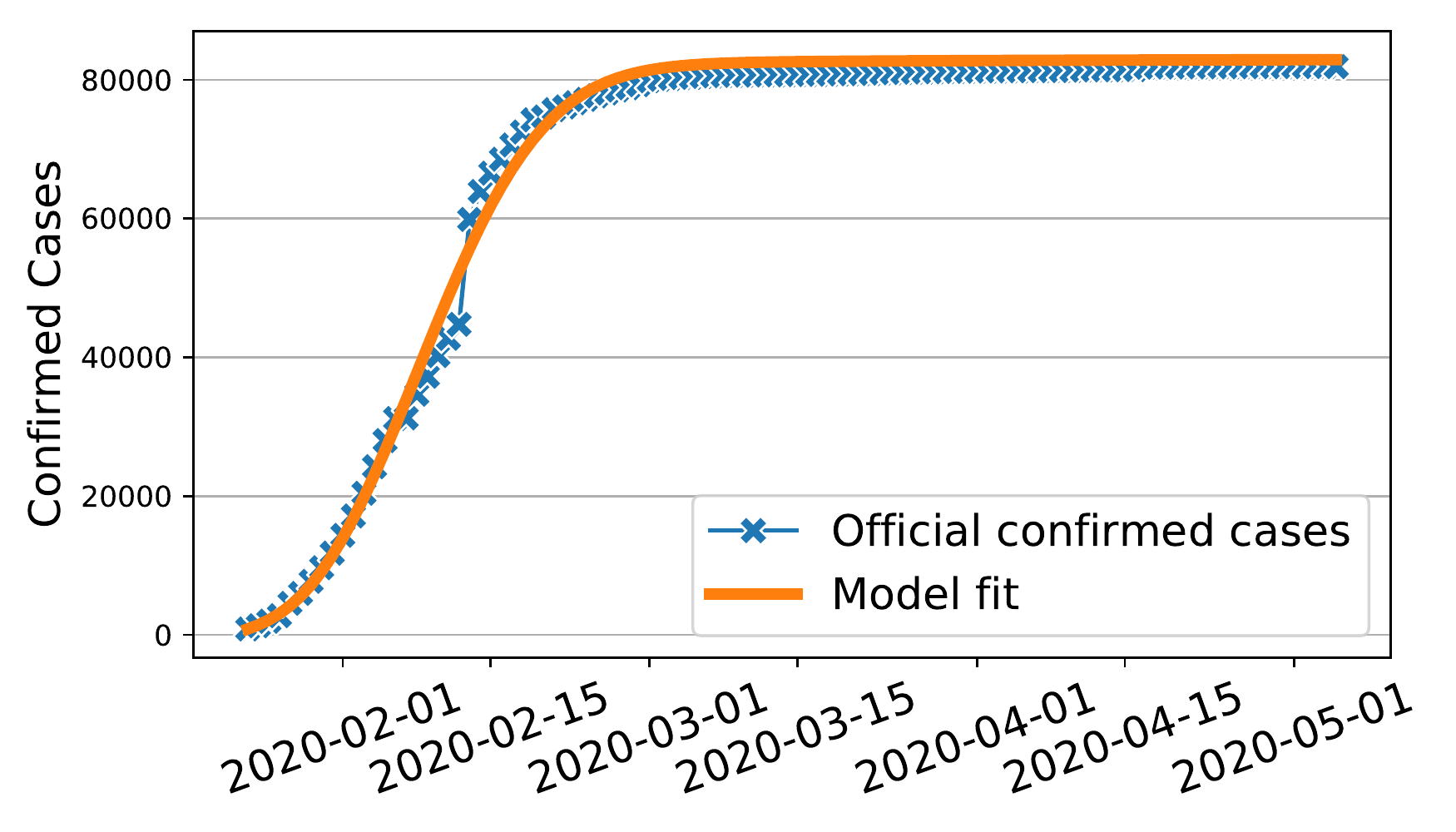}
    \caption{Comparison between official number of confirmed cases and fitting number from the SIR-X model on May 1, 2020 in China.}
    \label{fig:china-fit}
    \vspace{-5mm}
\end{figure}

Specifically, we use the uniform distribution as the parameter's prior distribution. 
And with the consideration of the nonlinear of SIR-X model, we adopt the Sequential Monte Carlo sampler to achieve the posterior distribution of model's parameters including the $\alpha$ and $\beta$.
Finally, we take the expected value of each parameter to construct the model.

In order to have stable results from the MCMC method, we use a priori conditions, i.e., $R_0 < R'_{0, free}$ and $\kappa_0 < \kappa$.
The results of MCMC methods may not be stable, i.e., the results of each execution may be different without a priori conditions. 
Thus, we introduce a priori conditions, i.e., $R_0 < R'_{0, free}$, $\kappa_0 < \kappa$ and the model fit number of accumulated confirmed cases should be equal or bigger than the official number of confirmed cases. $R'_{0, free}$ represents a maximum value of $R_0$.
We set $R'_{0, free}$ as 6.2, which is in accordance with the result from \cite{Maier2020} that the $R_0$ should be between 1.4 and 3.3. During the fitting process, if the a priori conditions are not met, the fitting process will be executed again until reaching a limit, e.g., 20 times of execution, in order to avoid infinite execution.
We assume that the quarantine measure is applied more strictly on the infected individuals than other public citizens, i.e., $\kappa_0 < \kappa$. 

In order to use the SIR-X model, we need to assume that few travelers and symptomatic infected individuals travel into or from a city. 
As reported in \cite{Xiong2020}, there is strong correlation between the inflows from Wuhan and the confirmed cases in a city. We assume that few infected individuals travelled into major cities after January 23, 2020 as Wuhan travel ban has been put in place since January 23, 2020 and few people went to other cities from Wuhan as shown in Figure \ref{fig:wuhan_migration}.
In addition, we assume that the number of infected individuals in the inflows from other cities can be ignored compared to the number of infected individuals among the local citizens in a city. 
With these two assumptions, we can use the SIR-X and MCMC to estimate parameters for each major cities in Mainland China based on the number of confirmed cases \footnote{COVID-19 statistics - \url{https://github.com/canghailan/Wuhan-2019-nCoV}} from January 23, 2020 to May 1, 2020.

\begin{figure}[ht]
\begin{subfigure}{.48\textwidth}
  \centering
  \includegraphics[width=.99\linewidth]{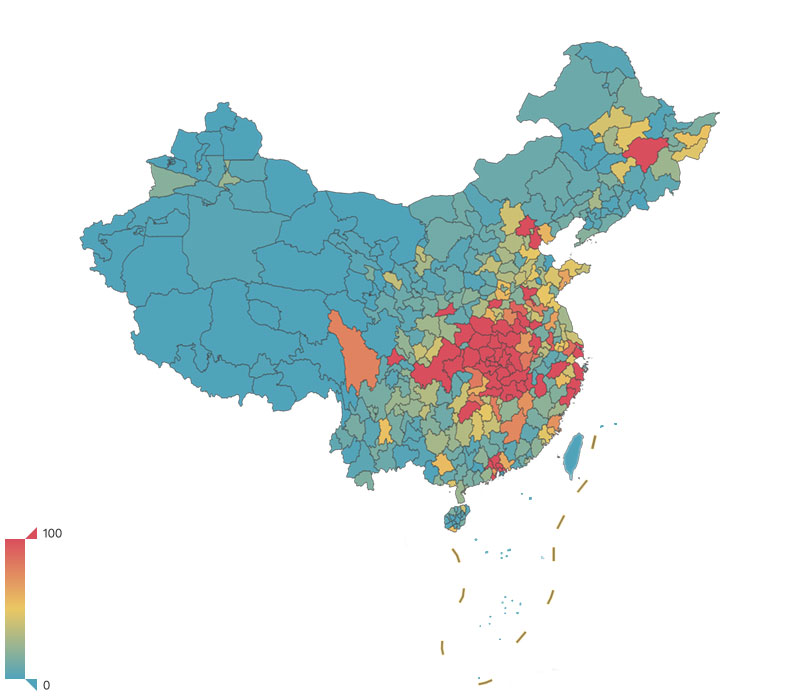}
  \caption{The official number (May 1, 2020) of confirmed cases.}
\end{subfigure}
\begin{subfigure}{.48\textwidth}
  \centering
  \includegraphics[width=.99\linewidth]{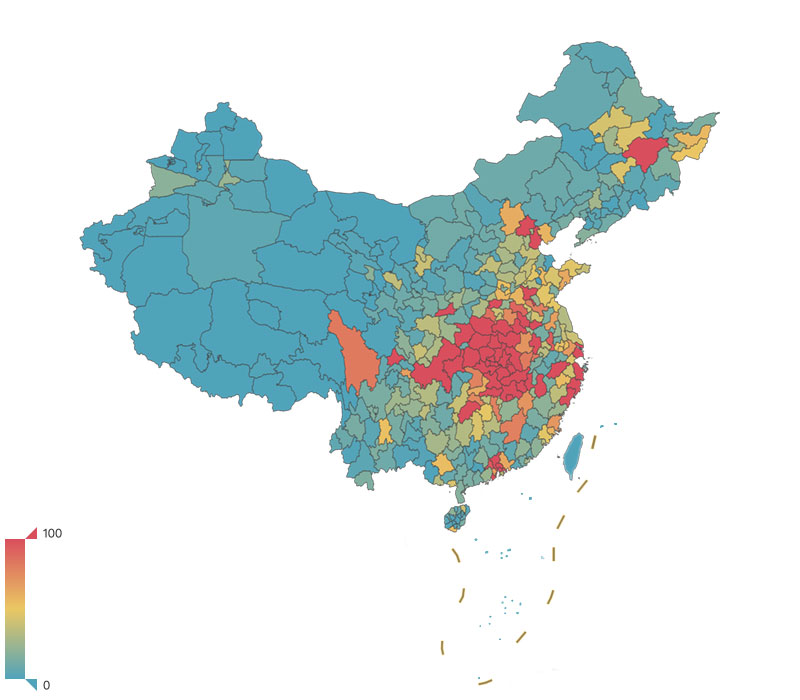}  
  \caption{The model fitting number (May 1, 2020) of confirmed cases.}
\end{subfigure}
\caption{The comparison between official number and model fitting number of confirmed cases in major cities of Mainland China.}
\label{fig:estimated-China}
\end{figure}

Figures \ref{fig:sub-beijing} - \ref{fig:sub-chongqing} illustrate the confirmed cases in several major cities of Mainland China. From figures, we can see that the combination of SIR-X and MCMC captures the number of confirmed cases in different cities very well, e.g., Beijing, Shanghai, Shenzhen, Wuhan and Chongqing.
However, the model does not well characterize the number of confirmed cases in Guangzhou as there are many (127 \footnote{Confirmed cases in Guangzhou from National Health Commission - \url{http://wjw.gz.gov.cn/ztzl/xxfyyqfk/yqtb/content/post_5815637.html}}) infected individuals from other countries, which cannot be captured by the SIR-X model.
In addition, we believe that the errors between the confirmed cases and fitted data are mainly due to the travelers from other countries in Beijing (174 confirmed cases from other countries) and Shanghai (326 confirmed cases from other countries).
Then, we calculate the confirmed cases of different provinces by adding the number of confirmed cases in each affiliated city. 
Figures \ref{fig:sub-Hubei} - \ref{fig:sub-Yunnan} shows the number of confirmed cases in several provinces of Mainland China. 
We can see that the SIR-X well captures the aggregated cases at province scale.
Furthermore, we use the same method to calculate the number of confirmed cases in Mainland China as shown in Figure \ref{fig:china-fit}.
Figure \ref{fig:estimated-China} shows that the combination of SIR-X model and MCMC captures well the number of confirmed cases at city scale. 


\begin{figure}[ht]
    \centering
    \includegraphics[width=0.75\textwidth]{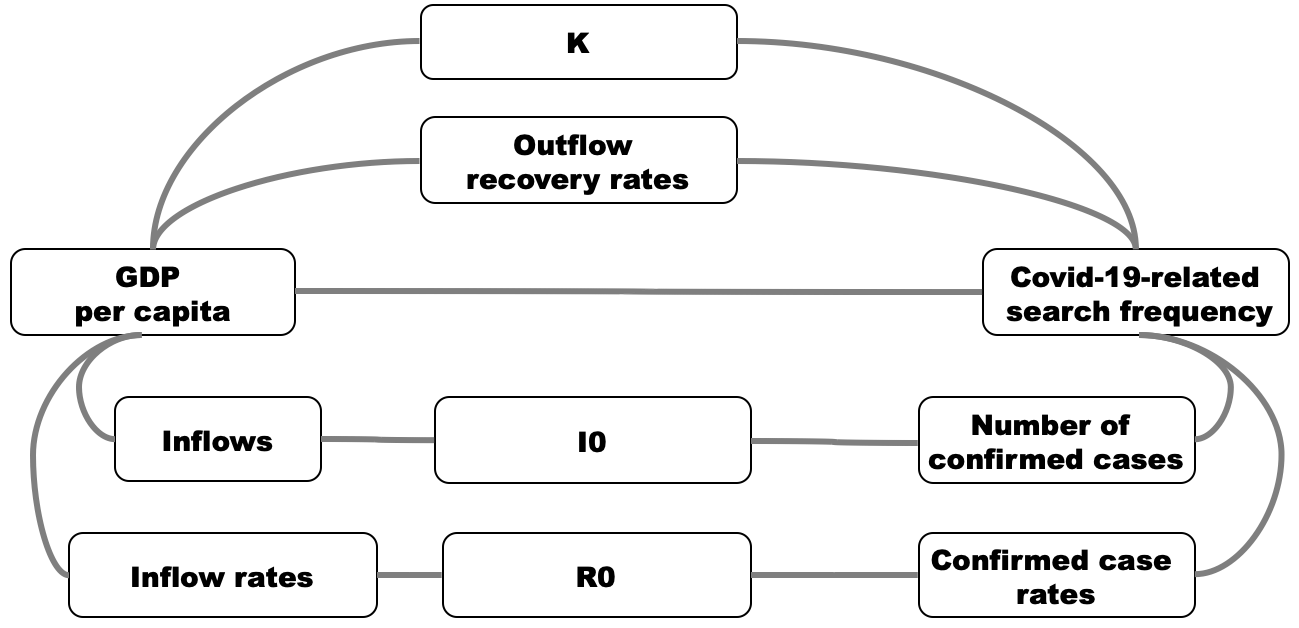}
    \caption{Significant correlation among different factors.}
    \label{fig:correlation}
    \vspace{-5mm}
\end{figure}

\begin{sidewaystable}[htbp]
\caption{Overall Results of Correlation Analysis}
\centering
\begin{tabular}{c c c}\hline
Correlations & Coeff. (R) & $p$-value  \\\hline
\multicolumn{3}{c}{\em Result I}\\\hline
GDP per capita vs. COVID-19-related search volume & $52.5\%$  & $<0.0001$  \\
GDP per capita vs. COVID-19-related search frequency & $63.5\%$  & $<0.0001$  \\ \hline
\multicolumn{3}{c}{Result II}\\\hline
GDP per capita vs. Inflows from Wuhan & $42.3\%$ & $<0.0001$ \\
Inflows from Wuhan vs. $I_0$ & $21.6\%$ & $<0.001$ \\
$I_0$ vs. Number of confirmed cases & $22.5\%$ & $<0.001$ \\
Number of confirmed cases vs. COVID-19-related search frequency \cite{Xiong2020} & $41.5\%$ & $<0.0001$ \\\hline
\multicolumn{3}{c}{Result III}\\\hline
GDP per capita vs. Inflows from Wuhan / population & $32.6\%$ & $<0.0001$ \\
Inflows from Wuhan / population vs. $R_0$ & $19.2\%$ & $<0.005$ \\
$R_0$ vs. Number of confirmed cases / population & $29.8\%$ & $<0.0001$ \\
Number of confirmed cases / population vs. COVID-19-related search frequency \cite{Xiong2020} & $21.4\%$ & $<0.001$ \\\hline
\multicolumn{3}{c}{\em Result IV}\\\hline
GDP per capita vs. $\kappa$ & $17.3\%$ & $<0.01$\\
COVID-19-related search frequency vs. $\kappa$ & $17.8\%$ & $<0.01$\\ \hline
\multicolumn{3}{c}{Result V}\\\hline
GDP per capita vs. Outflow recovery rates & $-46.5\%$ & $<0.0001$ \\
COVID-19-related search frequency vs. Outflow recovery rates & $-51.4\%$ & $<0.0001$ \\\hline
\end{tabular}
\label{tab:results}
\end{sidewaystable}

\section{Correlation study} 

Besides the number of confirmed cases (May 1, 2020), we collected three datasets, i.e., GDP, mobility and COVID-19-related search frequency for major cities in Mainland China. The GDP dataset was collected from \cite{ChineseStatics}, which characterized local economic development in 2019. The mobility data was captured from Baidu Maps and the COVID-19-related search frequency (the ratio between COVID-19-related search volume from January to March 2020 and population in each city) data was gathered from a widely-used Web search engine. We analyze the correlation among local economy, mobility, search behaviors and the parameters estimated based on the combination of the SIR-X model and the MCMC method as presented in Section \ref{sec:modeling}, for 238 cities in Mainland China (excluding Wuhan). 
We normalize inflow, outflow, search volume by the following formula:
\begin{equation}
    Normalize\mathrm{(data)} = \frac{\mathrm{data} - \mathrm{data}_{min}}{\mathrm{data}_{max}-\mathrm{data}_{min}}\ .
\end{equation}
In this way, the data in the study are curved into the range from 0 to 1 proportionally.
The results of our data-driven analysis are summarized in Table \ref{tab:results} and shown in Figure \ref{fig:correlation}. In this section, we present the observations obtained from the analysis.

\begin{figure}[ht]
    \centering
    \includegraphics[width=0.6\textwidth]{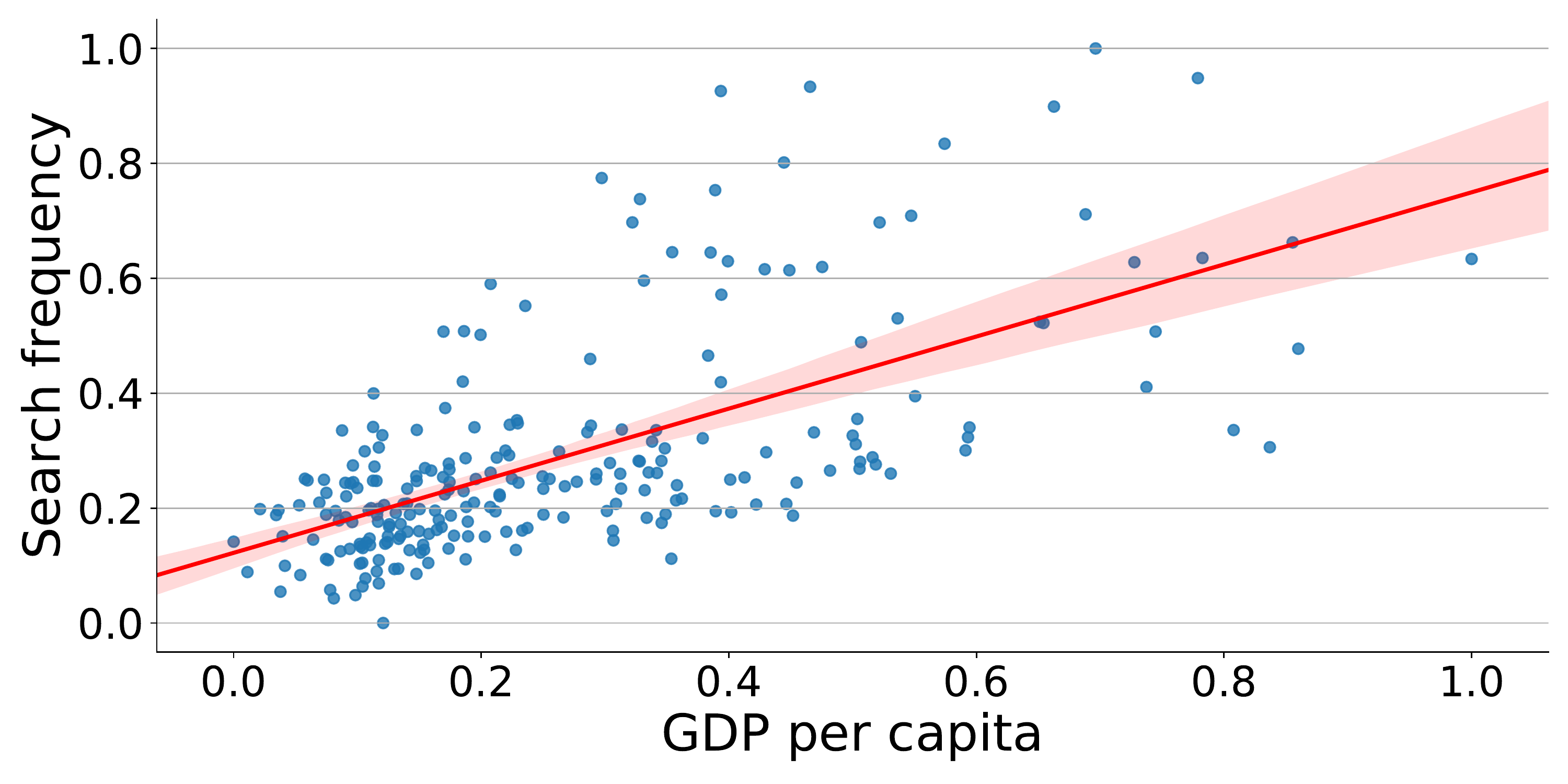}
    \caption{Significant positive correlation has been found between the GDP per capita and COVID-19-related search frequency for major cities in Mainland China. Shaded area represents the 95\% Confidence Interval (CI).}
    \label{fig:gdpp&searchp}
   \vspace{-4mm}
\end{figure}

\subsection{Significant positive correlations have been evidenced between local economy and COVID-19-related search frequency}

\emph{We have evidenced the significant positive correlations between local economy and COVID-19-related search frequency for major Chinese cities (Result I in Table \ref{tab:results}).} In order to analyze the correlation between two random variables, we calculated the Pearson correlation coefficients \cite{Benesty2009} and conducted the Student's T-test (two tails) to verify the significance test (the same for the following analysis in the paper).
\begin{figure}[ht]
  \centering
  \includegraphics[width=0.6\textwidth]{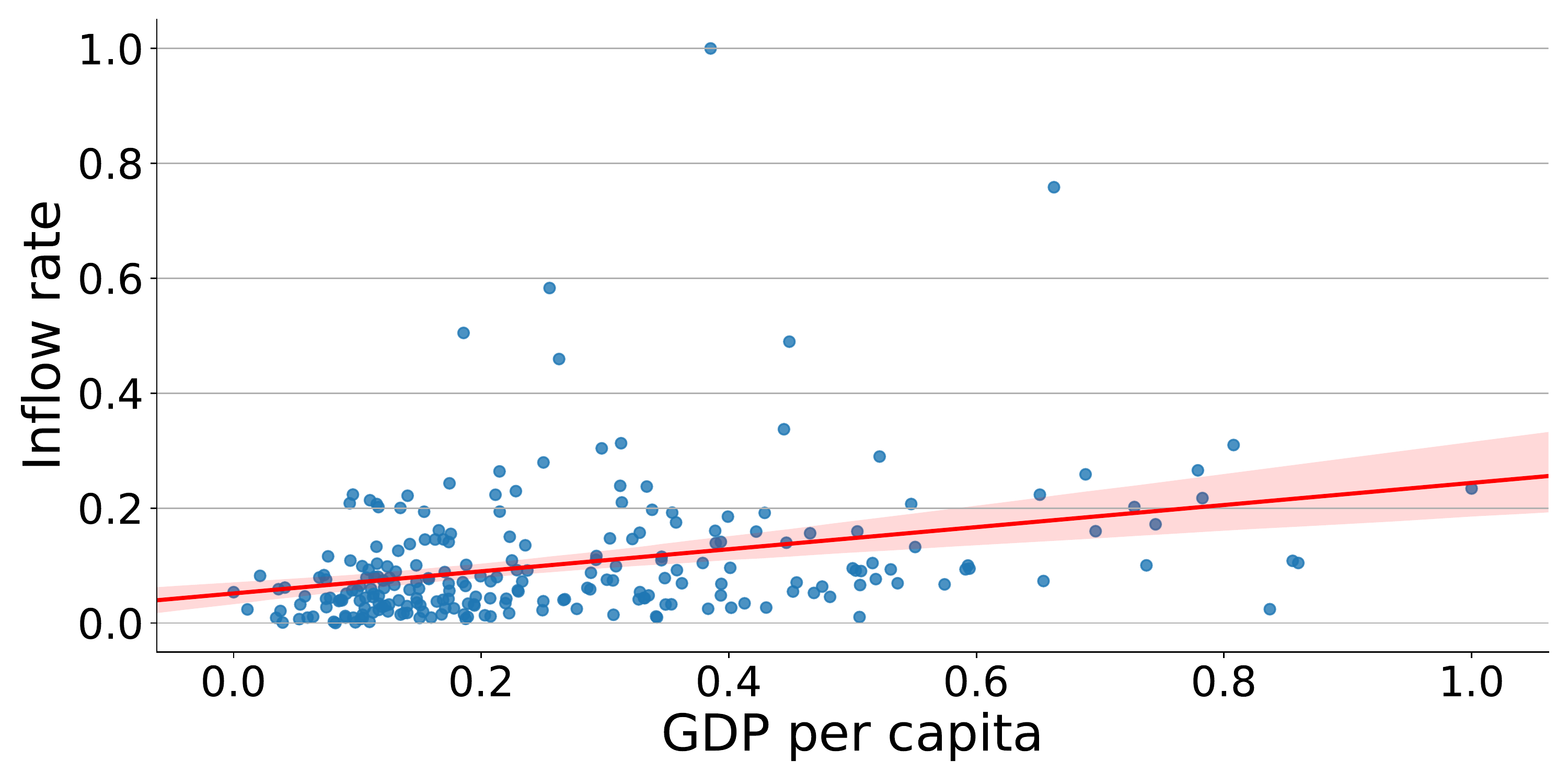}  
  \caption{Significant negative correlations have been found between the GDP per capita and inflow rate from Wuhan. Shaded area represents the 95\% CI.}
  \label{fig:gdpp&inflowsp}
  \vspace{-4mm}
\end{figure}
The Pearson correlation between the local GDP per capita and the total COVID-19-related search volume (between January and March 2020) is $R^{***}=52.5\%$ ($N=238$ and $p$-value$=3.06\times 10^{-18}< 0.0001$) for each city. However, we considered that this observation can be incurred by the scale of the city, as a larger city would have larger population, and would correspond to bigger GDP and high COVI-19-related search volume.

We therefore tested the significance of the correlations between GDP per capita and COVID-19-related search frequency, where we evidenced the significance in the correlations as $R^{***}=63.5\%$ ($N=238$ and $p$-value=$2.91\times10^{-28}< 0.0001$). 
In addition, in order to obviate the impact of the scale of the city, i.e., the impact of city population, we conducted partial correlation analysis \cite{Baba2004,Kenett2015} between GDP per capita and the COVID-19-related search frequency with the effects of the city population size (a controlling variable) removed. 
\begin{figure}[ht]
  \centering
  \includegraphics[width=0.6\textwidth]{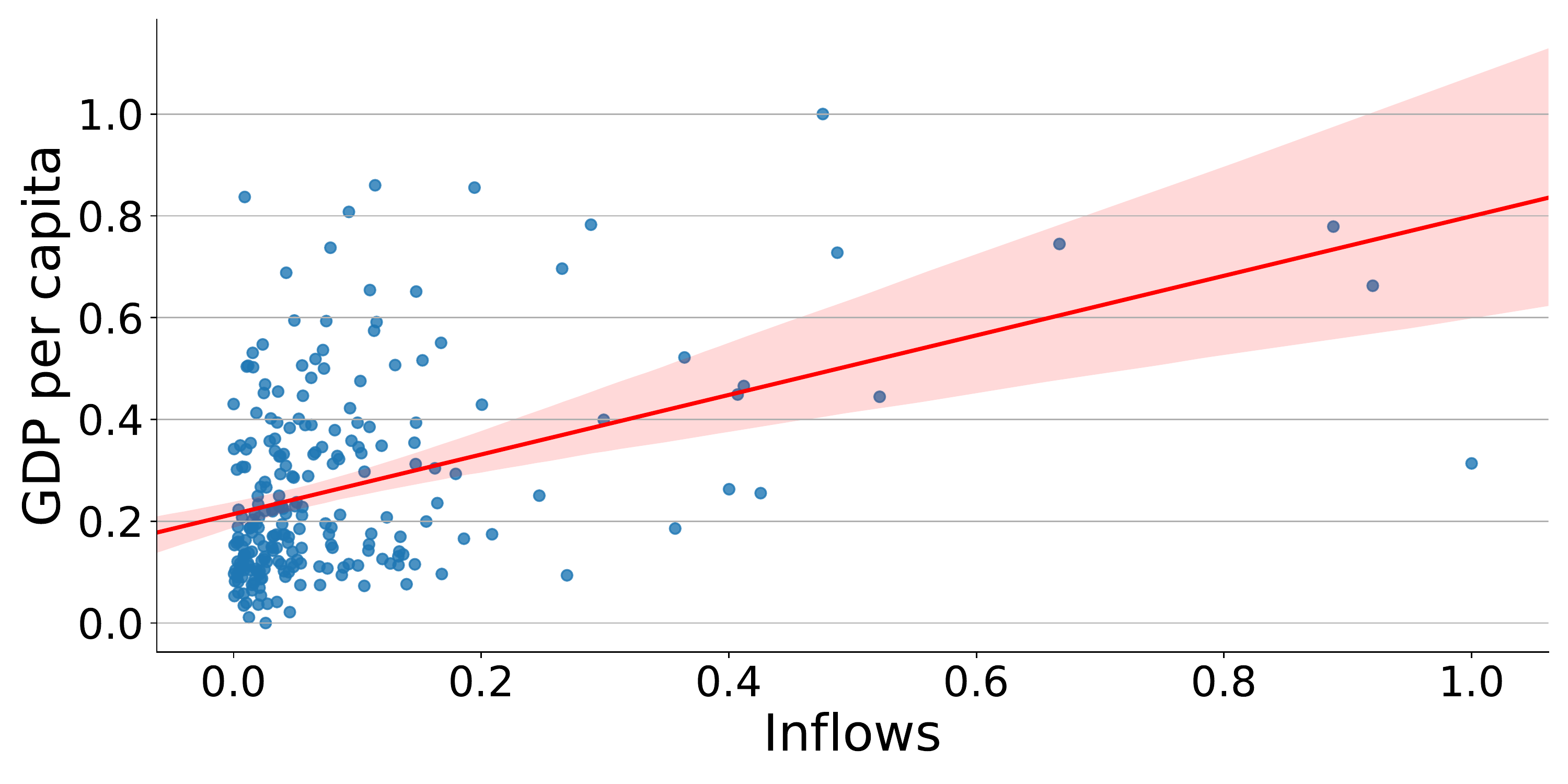} 
  \caption{Significant positive correlations have been found between the GDP per capita and inflows from Wuhan. Shaded area represents the 95\% CI.}
  \label{fig:gdpp&inflows}
  \vspace{-4mm}
\end{figure}
\begin{figure}[ht]
    \centering
    \includegraphics[width=0.6\textwidth]{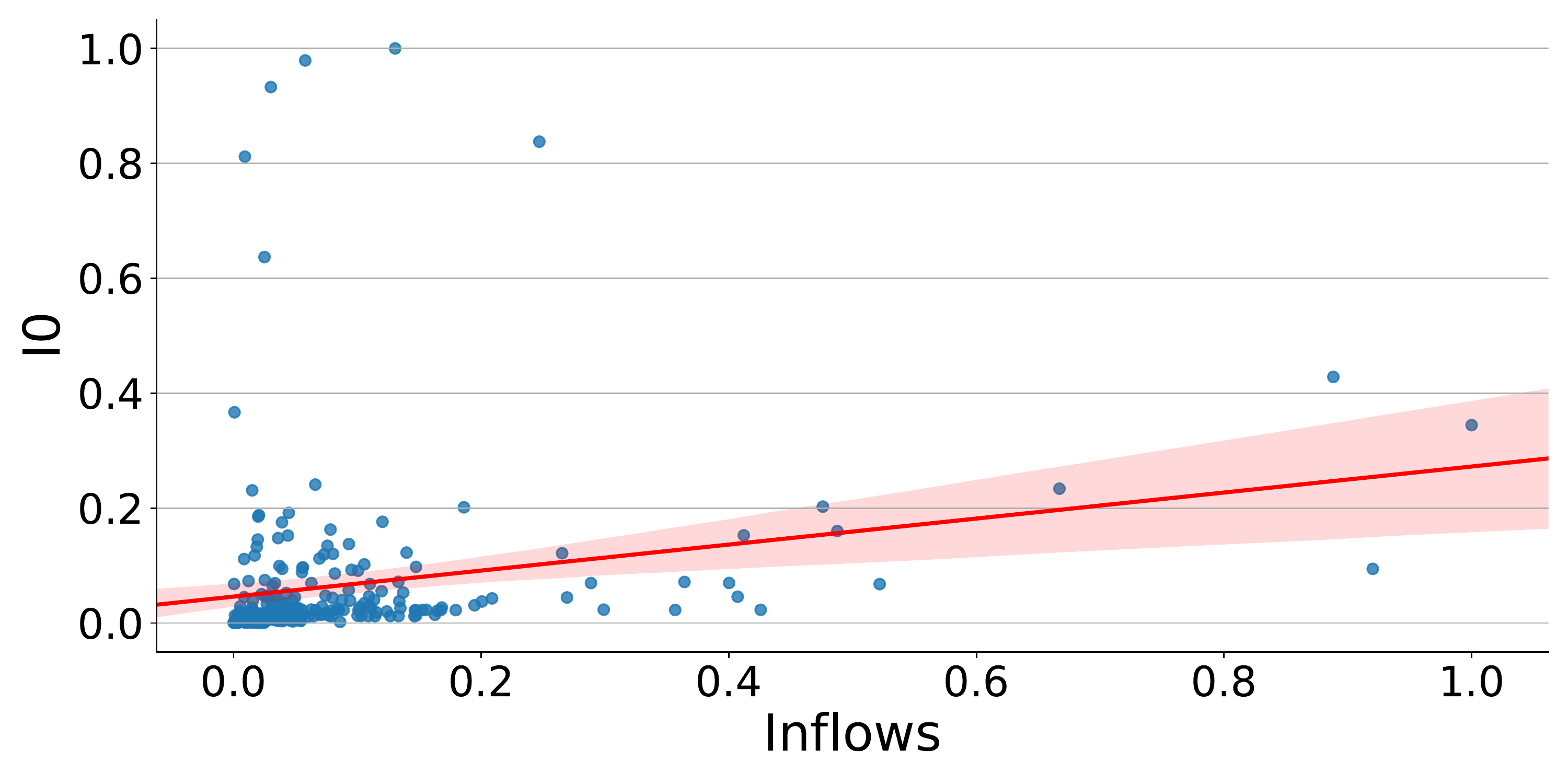}
    \caption{Significant positive correlations have been found between the inflows from Wuhan and $I_0$. Shaded area represents the 95\% CI.}
    \label{fig:inflows&I0}
   \vspace{-4mm}
\end{figure}
In order to estimate the partial correlation of random 
variables $X$ and $Y$ with the random variable $Z$ removed, we expressed the partial correlation coefficient in terms of the Pearson correlation coefficients as
\begin{equation}
    \rho (X,Y|Z) \equiv \frac{\rho (X,Y)-\rho (X,Z)\rho (Y,Z)}{\sqrt{(1-\rho ^2(X,Z))(1-\rho ^2(Y,Z))}} .
\end{equation}
We find a strong correlation with significance as well, such that $R^{***}=57.0\%$ ($N=238$ and $p$-value$=7.15\times 10^{-22}< 0.0001$). Thus, we can conclude that no matter whether the scale of the
city is big or small, the GDP per capita has a significant positive correlation with the COVID-19-related search frequency. Please see also in Figure~\ref{fig:gdpp&searchp} for the visualization of the correlations.

\subsection{Correlation analysis for the spread of the COVID-19 pandemic}

\emph{We have evidenced the significant positive correlation between the GDP per capita and inflows from Wuhan (Result II in Table \ref{tab:results}).} We hypothesized that cities with higher GDP per capita would attract larger inflows from Wuhan. Therefore, for every city in the study,  we correlated the GDP per capita and the inflows from Wuhan, where we obtained Pearson correlation coefficients of $R^{***}=42.3\%$ ($N=238$ and $p$-value$=9.88\times 10^{-12}< 0.0001$). In addition, in order to obviate the impact of the scale of the city, we correlated GDP per capita and the inflows rate from Wuhan, i.e., the ratio between inflows from Wuhan and the population. We found a strong positive correlation with significance as well, such that $R^{***}=32.6\%$ ($N=238$ and $p$-value$=2.67\times 10^{-7}< 0.0001$). Please see also in Figures~\ref{fig:gdpp&inflowsp} and \ref{fig:gdpp&inflows} for the visualization of the correlations. 

\begin{figure}[ht]
    \centering
    \includegraphics[width=0.6\textwidth]{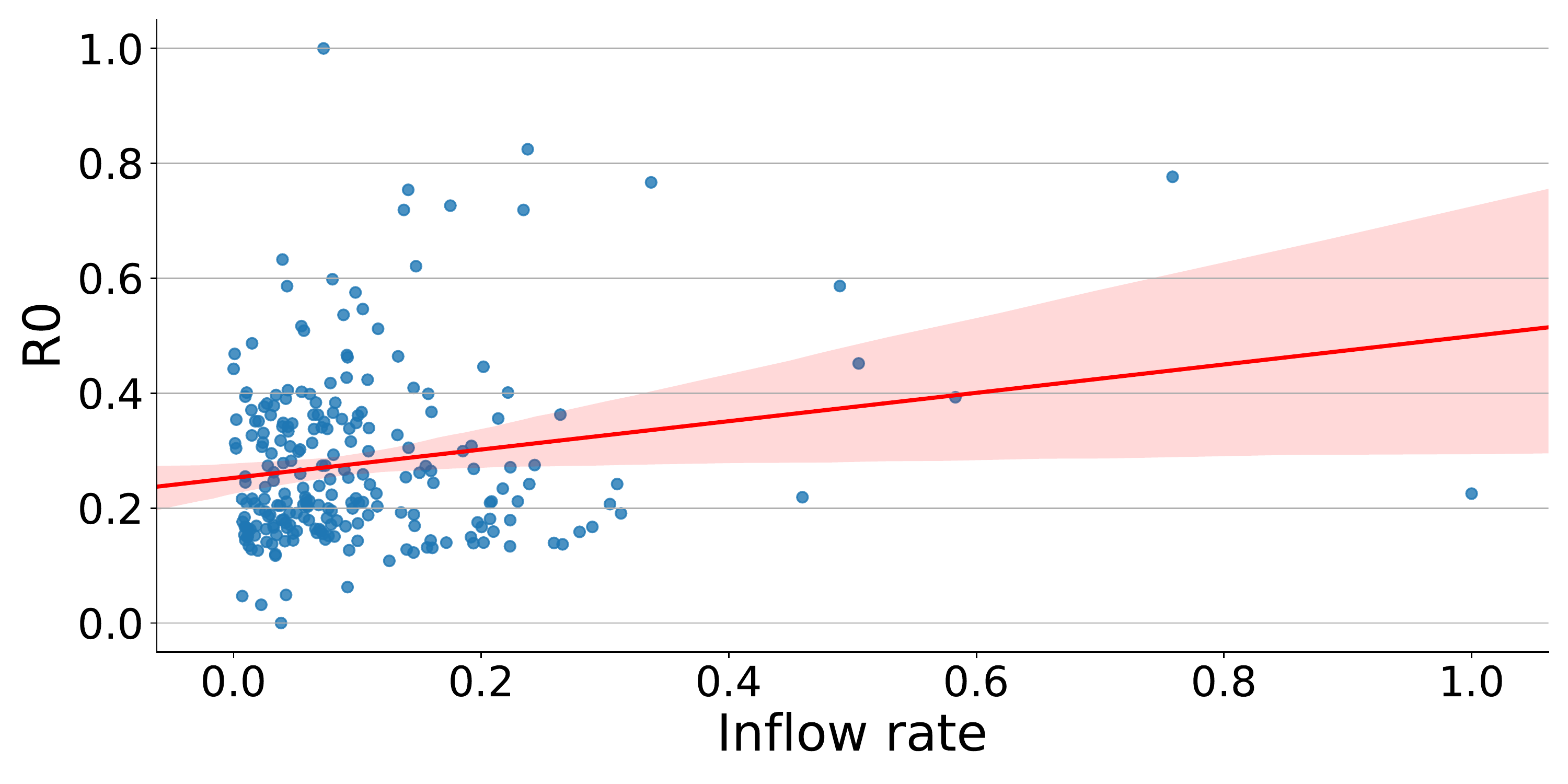}
    \caption{Significant negative correlations have been found between the inflow rate from Wuhan and $R_0$. Shaded area represents the 95\% CI.}
    \label{fig:inflowP&R0}
   \vspace{-4mm}
\end{figure}

\begin{figure}[ht]
    \centering
    \includegraphics[width=0.6\textwidth]{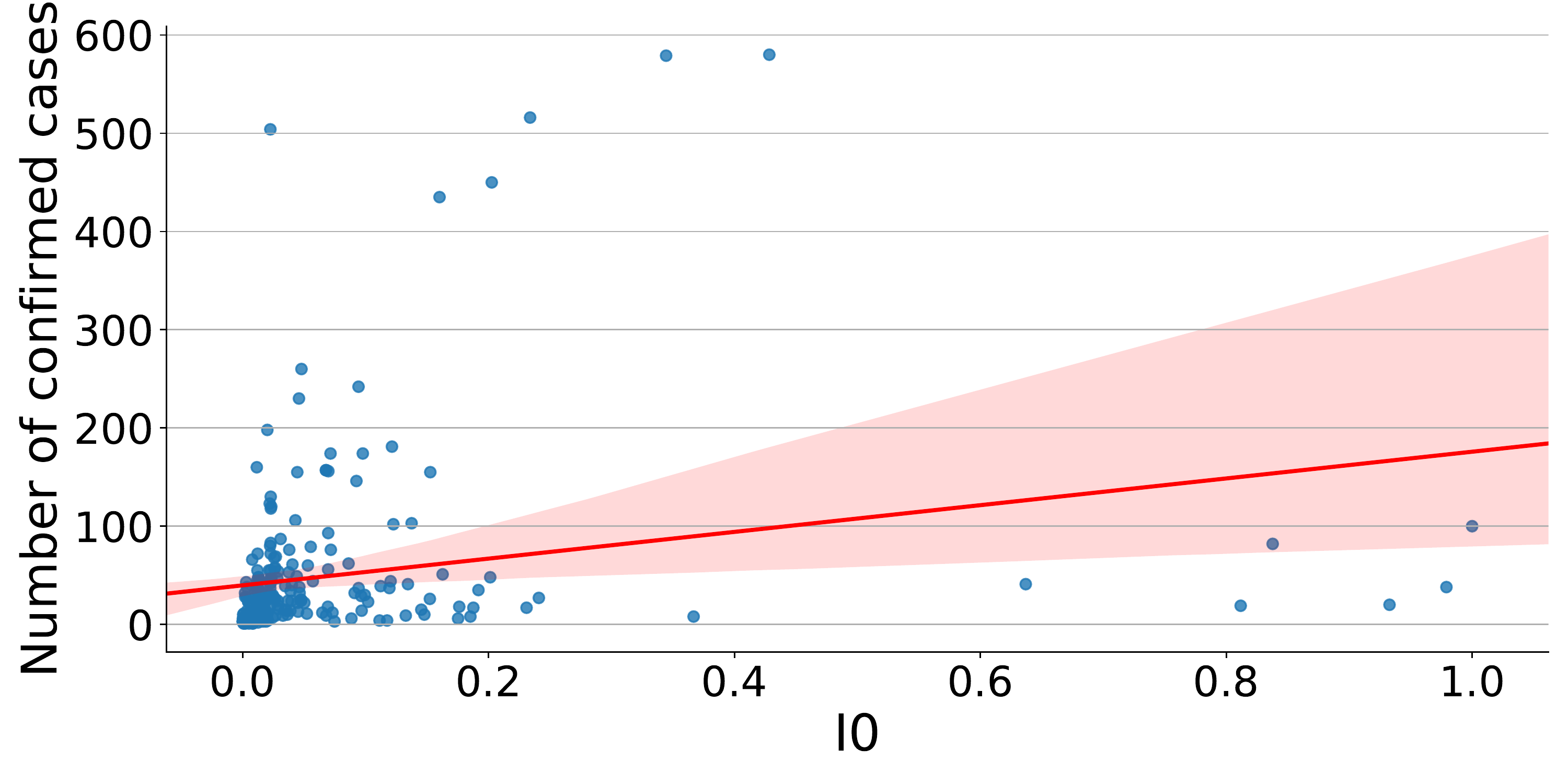}
    \caption{Significant positive correlations have been found between $I_0$ and the number of confirmed cases. Shaded area represents the 95\% CI.}
    \label{fig:I0&infections}
   \vspace{-4mm}
\end{figure}

\emph{We have evidenced the significant positive correlation between the inflows and $I_0$ (Result II in Table \ref{tab:results}) and the significant positive correlation between the inflow rate and $R_0$ (Result III in Table \ref{tab:results}).} We hypothesized that cities with larger inflows from Wuhan have more initial infected cases, i.e., $I_0$ in the SIR-X model. Thus, we correlated the inflows from Wuhan and $I_0$, where we obtained found a strong positive correlation with significance, such that $R^{**}=21.6\%$ ($N=238$ and $p$-value$=8.11\times 10^{-4}< 0.001$). In addition, we analyzed the correlation between the inflow rate and $R_0$, where we found a strong positive correlation with significance, such that $R^{*}=19.2\%$ ($N=238$ and $p$-value$=2.93\times 10^{-3}< 0.01$). Please see also in Figures~\ref{fig:inflows&I0} and \ref{fig:inflowP&R0} for the visualization of the correlations. 

\begin{figure}[ht]
    \centering
    \includegraphics[width=0.6\textwidth]{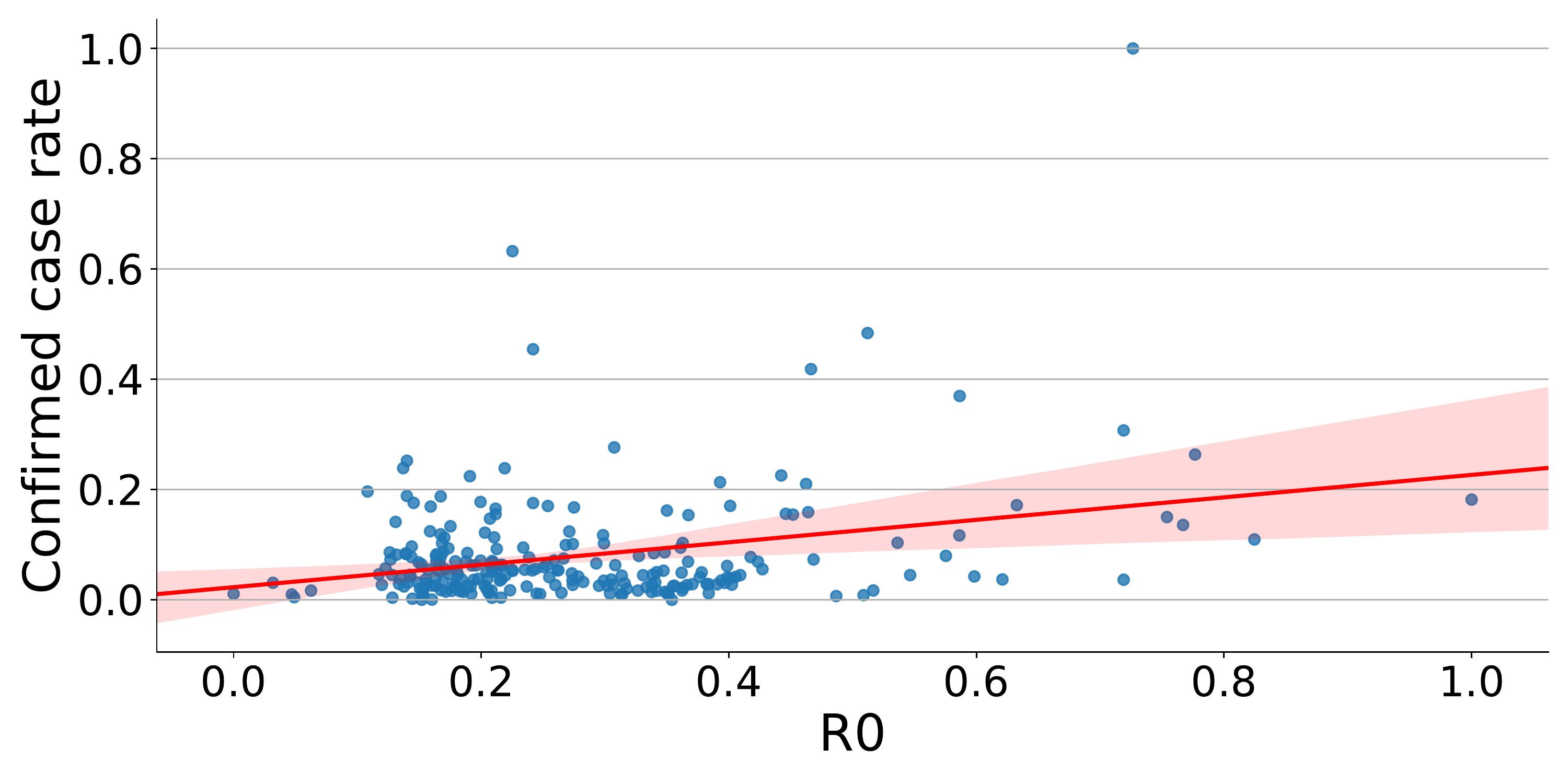}
    \caption{Significant negative correlations have been found between $R_0$ and confirmed case rate. Shaded area represents the 95\% CI.}
    \label{fig:R0&infectionP}
   \vspace{-4mm}
\end{figure}

\emph{We have evidenced the significance of the positive correlation between the number of initial infected individuals and the number of confirmed cases (Result II in Table \ref{tab:results}) and the positive correlation between $R_0$ and the number of confirmed case rate (Result III in Table \ref{tab:results}).} We hypothesized that cities with bigger $I_0$ finally have more confirmed cases. To this end, we performed correlation using $I_0$ and the number of confirmed cases on May 1, 2020. We found a significant positive correlation, such that $R^{**}=22.5\%$ with $N=238$ and $p$-value$=4.67\times 10^{-4}< 0.001$. In addition, we hypothesized that cities with bigger $R_0$ have more confirmed case rate, i.e., the ratio between the confirmed cases and the population. We performed correlation between $R_0$ and the confirmed case rate, where we obtained a significant positive correlation, such that $R^{***}=29.8\%$ with $N=238$ and $p$-value$=2.82\times 10^{-6}< 0.0001$. Please see also in Figures~\ref{fig:I0&infections} and \ref{fig:R0&infectionP} for the visualization of the correlations. 


\begin{figure}[ht]
    \centering
    \includegraphics[width=0.6\textwidth]{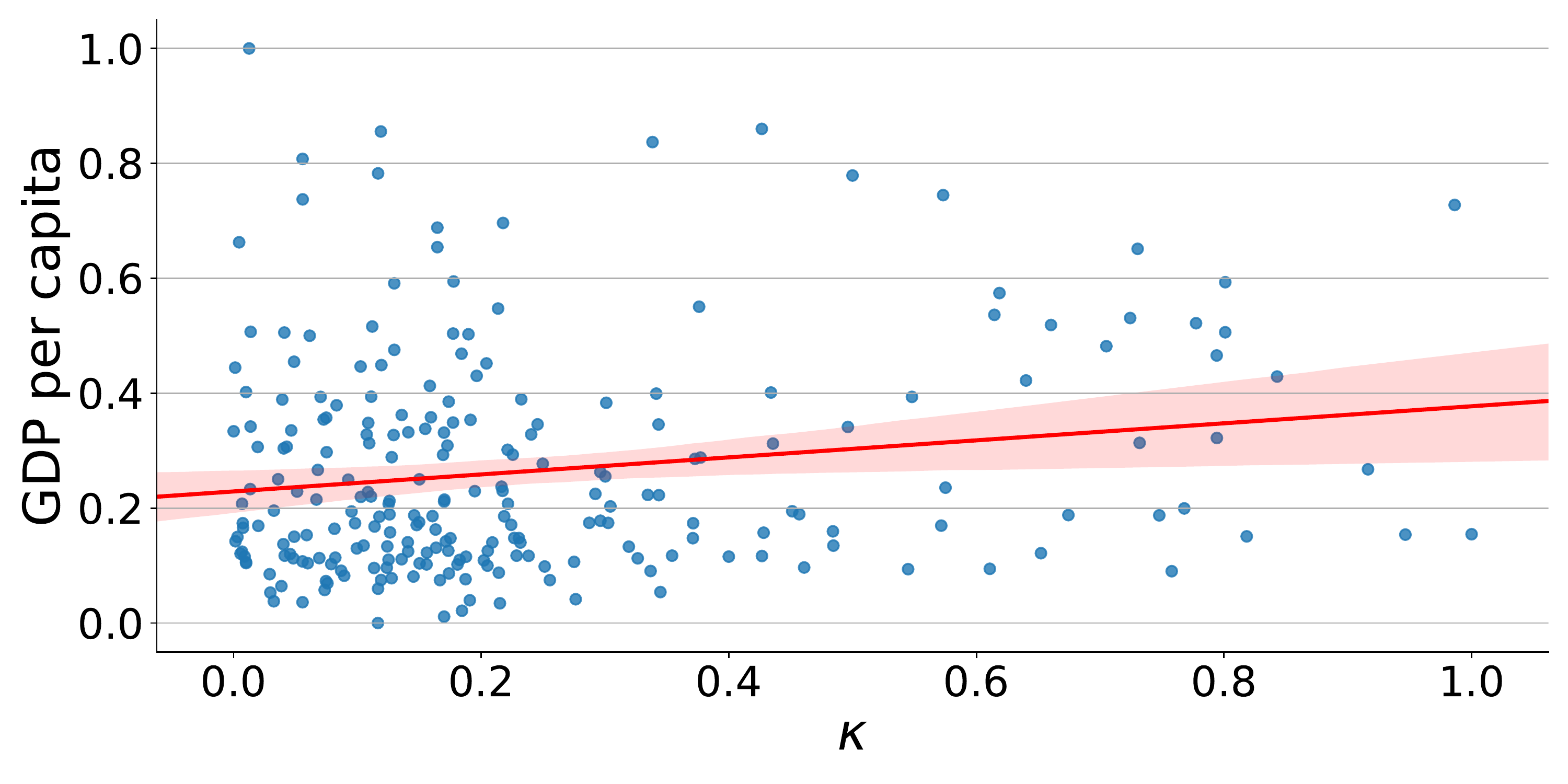}
    \caption{Significant positive correlations have been found between the GDP per capita and $\kappa$. Shaded area represents the 95\% CI.}
    \label{fig:gdpp&kappa}
   \vspace{-4mm}
\end{figure}

We obtained a strong positive correlation with significance between the number of confirmed cases and the COVID-19-related search frequency, such that $R^{***}=41.5\%$ ($N=238$ and $p$-value$=2.45\times 10^{-11}< 0.0001$). 
Furthermore, we obtained a strong positive correlation with significance between confirmed case rate and the COVID-19-related search frequency, such that $R^{**}=21.4\%$ ($N=238$ and $p$-value$=9.01\times 10^{-4}< 0.001$). Similar results are also reported in \cite{Xiong2020}.



We thus can conclude that for each major city in the study, the GDP per capita and the factors that incur infections, e.g., inflows from Wuhan, $I_0$, $R_0$, have significant positive correlation. This indicates that the rich cities attract more inflows from Wuhan, which caused infections and in order to fight against COVID-19, the citizens in the rich cities tend to perform more search activities in order to be well-informed.

\subsection{Correlation analysis for the execution of containment measures}

In this section, we analyze the correlation among local economy, information acquisition and containment measures. 
As the quarantine measure (see details in Section \ref{sec:modeling}) is directly related to the number of confirmed cases, we analyze the correlation between $\kappa$ and other factors (local economy and information acquisition). In addition, we are also interested in the realization of inter-city containment measures, i.e., the outflow recovery rate (the ratio between the outflows of 2020 and that of 2019). 

\begin{figure}[ht]
    \centering
    \includegraphics[width=0.6\textwidth]{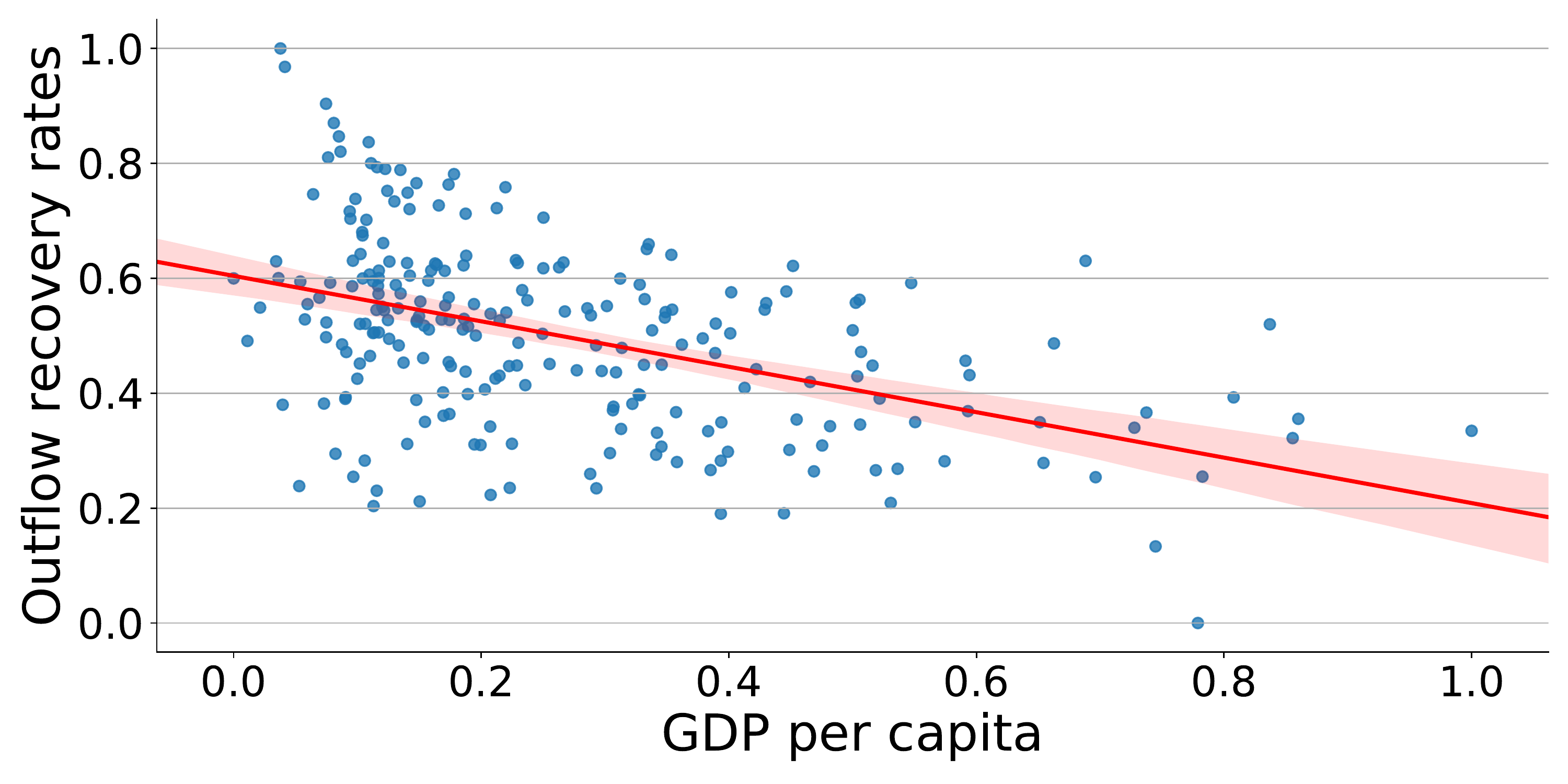}
    \caption{Significant negative correlations have been found between the GDP per capita and the outflow recovery rate. Shaded area represents the 95\% CI.}
    \label{fig:gdpp&recovery}
   \vspace{-4mm}
\end{figure}

\begin{figure}[ht]
    \centering
    \includegraphics[width=0.6\textwidth]{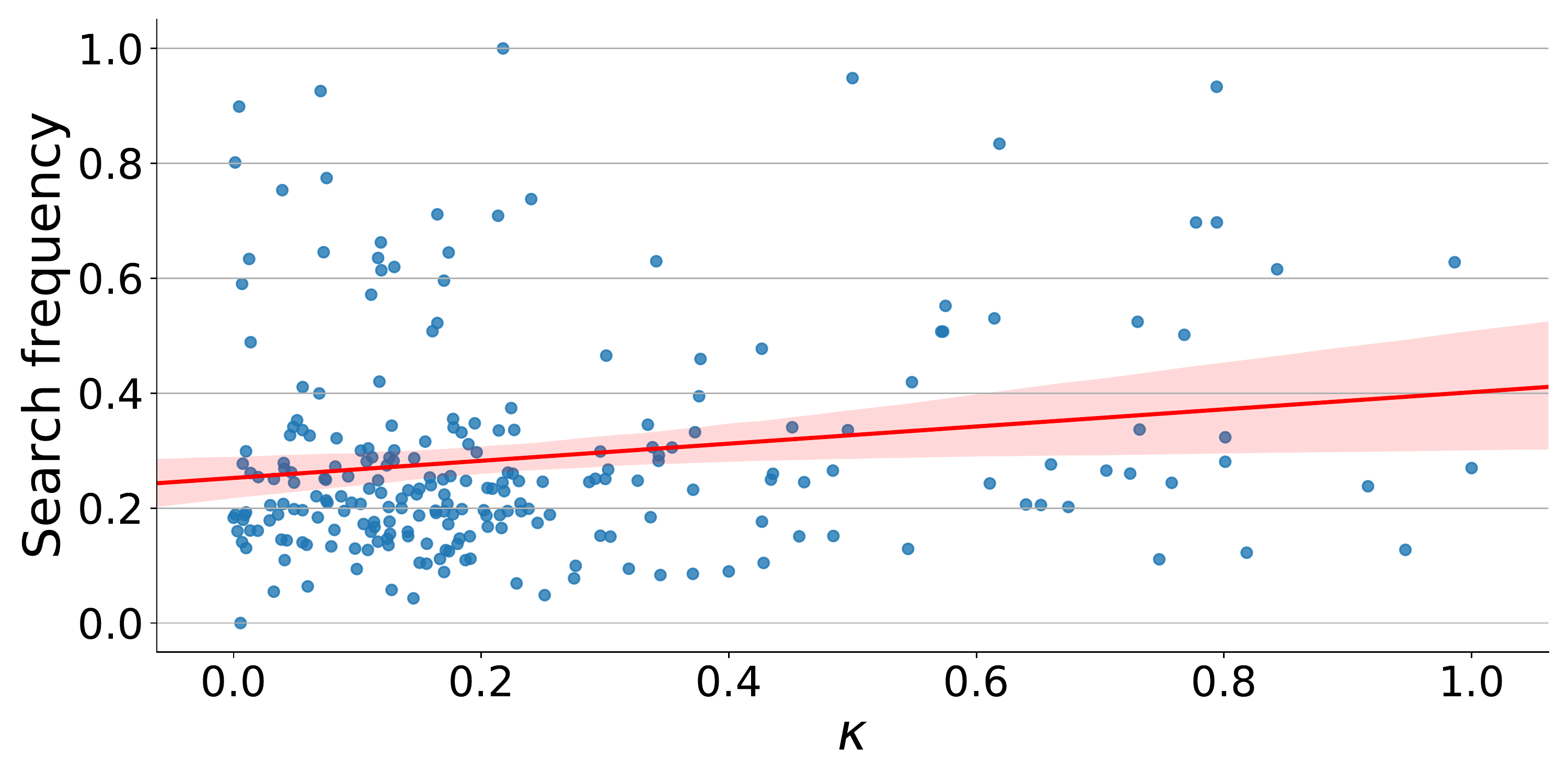}
    \caption{Significant positive correlations have been found between the COVID-19-related search frequency and $\kappa$. Shaded area represents the 95\% CI.}
    \label{fig:search&kappa}
   \vspace{-4mm}
\end{figure}

\emph{We have evidenced the significance of the positive correlation between the execution of the quarantine measure and GDP per capita for major Chinese cities in the study (Result IV in Table \ref{tab:results}).} 
Among all 238 cities in the correlation study, we hypothesized that people with high GDP per capita would try harder to realize the quarantine measure for infected individuals. Therefore, we correlated the GDP per capita and $\kappa$, where Pearson correlation coefficients are $R^{*}=17.3\%$ ($N=238$ and $p$-value$=7.46\times 10^{-3}< 0.01$). 
Furthermore, we correlated the GDP per capita and the outflow recovery rate, where we obtained Pearson correlation coefficients of $R^{***}=-46.5\%$ ($N=238$ and $p$-value$=3.82\times 10^{-14}< 0.0001$). 
The correlation analysis result suggests that people with higher GDP per capita are more likely to apply the quarantine measure.
Please see also in Figures~\ref{fig:gdpp&kappa} and \ref{fig:gdpp&recovery} for the visualization of the correlations. 

\emph{We have evidenced the significance of the positive correlations between the realization of quarantine measure and the COVID-19-related search frequency.} We performed correlation using the COVID-19-related search frequency and $\kappa$. We found a significant positive correlation, such that $R^{*}=17.8\%$ with $N=238$ and $p$-value$=5.84\times 10^{-3}< 0.01$. 
Please see also in Figures~\ref{fig:search&kappa} for the visualization of the correlations. 
We found negative correlations between the COVID-19-related search frequency and the outflow recovery rate are stronger with $R^{***}=-51.4\%$ ($N=238$ and $p$-value$=1.88\times 10^{-17}< 0.0001$) (similar result is also reported in \cite{Xiong2020}). The correlation analysis result suggests that people with more per capita COVID-19-related search frequency are more likely to apply the containment measures, i.e., separate the infected individuals and small outflow recovery rate.

 
We can conclude that for every city in the study the GDP per capita and the COVID-19-related search frequency have significant positive correlation to the realization of containment measures. 
We believe it is due to the will to avoid the risk to be infected and the natural response to the fear and massive panics~\cite{ren2020fear, Xiong2020}. In addition, the reason also goes to the fact that the people in the cities of higher GDP per capita tend to have bigger capacity or more tolerance to apply the containment measures.

\section{Conclusion}
In this work, we first exploit the SIR-X model and MCMC method to estimate the parameters related to the COVID-19 pandemic at the scale of city. Then, we examined the correlation between the local economy and the spread of COVID-19 pandemic and the execution of containment measures in major cities of Mainland China. We conducted correlation analysis based on the mobility data and search data from Baidu Maps and Baidu Search Engine in Mainland China. Our analysis brings novel knowledge of the correlation among different factors related to the COVID-19 pandemic. The cities of higher GDP per capita attracts bigger inflows from Wuhan, which cause more confirmed cases. However, the demands of information from individuals becomes higher, which incurs the reaction to apply the containment measures. Furthermore, well-informed individuals are more likely to apply the intra- and inter- city containment measures, i.e., quarantine of infected individuals and reducing going to other cities. The implications of these correlations include that, the better the local economy is and the more that timely information acquisition is attained by residents, the better the containment measures are realized, which help fight against the COVID-19 pandemic in major cities of Mainland China.

\section*{Acknowledgement}
To protect Baidu users' data privacy, all experiments in this paper were carried out using anonymous data and secure data analytics provided by Baidu Data Federation Platform (Baidu FedCube). For data accesses and usages, please contact us via \{fedcube, shubang\}@baidu.com.

\section*{Author contributions statement}
J. Liu formulated the research problems and drafted the manuscript. J. Liu, X. Wang and J. Huang collected data from Baidu, conducted the experiments, and performed data analysis. S. Huang and  H. An collected the consensus data and carried out the data visualization. X. Wang, J. Huang, H. Xiong and D. Dou revised the whole paper. D. Dou and H. Wang proposed the research, coordinated the research efforts, and oversaw the whole research process.

\bibliographystyle{plos2015}
\bibliography{main}

\begin{thebibliography}{10}

\bibitem{chinazzi2020effect}
Ahinazzi M, Davis JT, Ajelli M, Gioannini C, Litvinova M, Merler S, et~al.
\newblock The effect of travel restrictions on the spread of the 2019 novel
  coronavirus ({COVID-19}) outbreak.
\newblock Science. 2020;368(6489):395--400.
\newblock doi:{10.1126/science.aba9757}.

\bibitem{JH2020}
University JH. {Johns Hopkins University Coronavirus Resource Center}; 2020.
\newblock \url{http://https://coronavirus.jhu.edu/}.

\bibitem{Deslandes2020}
Deslandes A, Berti V, Tandjaoui-Lambotte Y, Alloui C, Carbonnelle E, Zahar JR,
  et~al.
\newblock {SARS-CoV-2} was already spreading in France in late December 2019;
  2020.

\bibitem{novel2020epidemiological}
{The Novel Coronavirus Pneumonia Emergency Response Epidemiology Team}.
\newblock The epidemiological characteristics of an outbreak of 2019 novel
  coronavirus diseases ({COVID-19}) in {China}.
\newblock {China} CDC Weekly. 2020;41(2):145.

\bibitem{sohrabi2020world}
Sohrabi C, Alsafi Z, O’Neill N, Khan M, Kerwan A, Al-Jabir A, et~al.
\newblock {World Health Organization declares global emergency: A review of the
  2019 novel coronavirus ({COVID-19})}.
\newblock International Journal of Surgery. 2020;76:71--76.

\bibitem{kraemer2020effect}
Kraemer MU, Yang CH, Gutierrez B, Wu CH, Klein B, Pigott DM, et~al.
\newblock The effect of human mobility and control measures on the {COVID-19}
  epidemic in {China}.
\newblock Science. 2020;368(6490):493--497.
\newblock doi:{10.1126/science.abb4218}.

\bibitem{tian2020investigation}
Tian H, Liu Y, Li Y, Wu CH, Chen B, Kraemer MU, et~al.
\newblock An investigation of transmission control measures during the first 50
  days of the {COVID-19} epidemic in {China}; 2020.

\bibitem{ferretti2020quantifying}
Ferretti L, Wymant C, Kendall M, Zhao L, Nurtay A, Abeler-D{\"o}rner L, et~al.
\newblock Quantifying {SARS-COV-2} transmission suggests epidemic control with
  digital contact tracing; 2020.

\bibitem{Engle2020}
Engle S, Stromme J, Zhou A.
\newblock {Staying at Home: Mobility Effects of COVID-19}; 2020.

\bibitem{Remuzzi2020}
Remuzzi A, Remuzzi G.
\newblock COVID-19 and Italy: what next?
\newblock The Lancet. 2020;395(10231):1225 -- 1228.

\bibitem{Maier2020}
Maier BF, Brockmann D.
\newblock Effective containment explains subexponential growth in recent
  confirmed COVID-19 cases in China.
\newblock Science. 2020;368(6492):742--746.
\newblock doi:{10.1126/science.abb4557}.

\bibitem{Chen2020}
Chen S, Yang J, Yang W, Wang C, B\"{a}rnighausen T.
\newblock {COVID-19} control in China during mass population movements at New
  Year.
\newblock Lancet. 2020;395(10226):764--766.
\newblock doi:{10.1016/S0140-6736(20)30421-9}.

\bibitem{ren2020fear}
Ren SY, Gao RD, Chen YL.
\newblock Fear can be more harmful than the severe acute respiratory syndrome
  coronavirus 2 in controlling the corona virus disease 2019 epidemic.
\newblock World Journal of Clinical Cases. 2020;8(4):652.

\bibitem{yang2020modified}
Yang Z, Zeng Z, Wang K, Wong SS, Liang W, Zanin M, et~al.
\newblock {Modified SEIR and AI prediction of the epidemics trend of {COVID-19}
  in {China} under public health interventions}.
\newblock Journal of Thoracic Disease. 2020;12(3):165.

\bibitem{li2020novel}
Li H, Chen X, Huang H.
\newblock The novel coronavirus outbreak: what can be learned from {China} in
  public reporting?
\newblock Global Health Research and Policy. 2020;5(1):1--3.

\bibitem{chen2020covid}
Chen S, Yang J, Yang W, Wang C, B{\"a}rnighausen T.
\newblock {COVID-19} control in {China} during mass population movements at New
  Year.
\newblock The Lancet. 2020;395(10226):764--766.

\bibitem{zhang2020challenges}
Zhang J, Lin G, Zeng J, Lin J, Tian J, Li G.
\newblock Challenges of {SARS-COV-2} and lessons learnt from {SARS} in
  Guangdong Province, {China}.
\newblock Journal of Clinical Virology. 2020;126:104341.

\bibitem{chen2020correlation}
Chen H, Chen Y, Sun B, Wang P, Wen L, Lian Z, et~al.
\newblock {Correlation between the migration scale index and the number of new
  confirmed Novel Coronavirus Pneumonia cases in {China}}.
\newblock Virus. 2020;11:3.

\bibitem{zhong2020correlation}
Zhong P, Guo S, Chen T.
\newblock {Correlation between travellers departing from Wuhan before the
  Spring Festival and subsequent spread of {COVID-19} to all provinces in
  {China}}.
\newblock Journal of Travel Medicine. 2020;27(3):1--4.

\bibitem{Xiong2020}
Xiong H, Liu J, Huang J, Huang S, An H, Kang Q, et~al.
\newblock Understanding the Collective Responses of Populations to the COVID-19
  Pandemic in Mainland China; 2020.

\bibitem{Gilks2005}
Gilks WR.
\newblock Markov Chain Monte Carlo; 2005.

\bibitem{huang2020quantifying}
Huang J, Wang H, Xiong H, Fan M, Zhuo A, Li Y, et~al.. Quantifying the Economic
  Impact of COVID-19 in Mainland China Using Human Mobility Data; 2020.

\bibitem{Atkeson2020}
Atkeson A.
\newblock {What Will Be the Economic Impact of COVID-19 in the US? Rough
  Estimates of Disease Scenarios}.
\newblock National Bureau of Economic Research; 2020. 26867.

\bibitem{gao2014quantifying}
Gao L, Song C, Gao Z, Barab{\'a}si AL, Bagrow JP, Wang D.
\newblock Quantifying information flow during emergencies.
\newblock Scientific reports. 2014;4:3997.

\bibitem{Toda2020}
Toda AA.
\newblock {Susceptible-Infected-Recovered (SIR) Dynamics of COVID-19 and
  Economic Impact}; 2020.

\bibitem{Simha2020}
Simha A, Prasad RV, Narayana S.
\newblock {A simple Stochastic SIR model for COVID 19 Infection Dynamics for
  Karnataka: Learning from Europe}; 2020.

\bibitem{Peng2020}
Peng L, Yang W, Zhang D, Zhuge C, Hong L.
\newblock Epidemic analysis of COVID-19 in China by dynamical modeling; 2020.

\bibitem{Hou2020}
Hou C, Chen J, Zhou Y, Hua L, Yuan J, He S, et~al.
\newblock The effectiveness of quarantine of Wuhan city against the Corona
  Virus Disease 2019 ({COVID}-19): A well-mixed SEIR model analysis; 2020.

\bibitem{Tang2020}
Tang S, Tang B, Bragazzi NL, Xia F, Li T, He S, et~al.
\newblock Analysis of COVID-19 epidemic traced data and stochastic discrete
  transmission dynamic model; 2020.

\bibitem{Yang2020}
Yang Z, Zeng Z, Wang K, Wong SS, Liang W, Zanin M, et~al.
\newblock Modified SEIR and AI prediction of the epidemics trend of COVID-19 in
  China under public health interventions.
\newblock Journal of Thoracic Disease. 2020;12(3):165--174.

\bibitem{Wu2020}
Wu JT, Leung K, Leung GM.
\newblock Nowcasting and Forecasting the Potential Domestic and International
  Spread of the 2019-nCoV Outbreak Originating in Wuhan, China: A Modelling
  Study.
\newblock Lancet. 2020;395(10225):689‐697.
\newblock doi:{10.1016/S0140-6736(20)30260-9}.

\bibitem{ChineseStatics}
of~Statistics~of China NB. {National Bureau of Statistics of China}; 2019.
\newblock \url{http://www.stats.gov.cn/}.

\bibitem{Benesty2009}
Benesty J, Chen J, Huang Y, Cohen I.
\newblock Pearson Correlation Coefficient.
\newblock Noise Reduction in Speech Processing. 2009;2:1--4.

\bibitem{Baba2004}
Baba K, Shibata R, Sibuya M.
\newblock Partial correlation and conditional correlation as measures of
  conditional independence.
\newblock Australian \& New Zealand Journal of Statistics. 2004;46(4):657--664.
\newblock doi:{10.1111/j.1467-842X.2004.00360.x}.

\bibitem{Kenett2015}
Kenett DY, Huang X, Vodenska I, Havlin S, Stanley HE.
\newblock Partial correlation analysis: applications for financial markets.
\newblock Quantitative Finance. 2015;15(4):569--578.
\newblock doi:{10.1080/14697688.2014.946660}.

\end{thebibliography}




\end{document}